\documentclass[11pt]{article} 

\usepackage{graphicx} 
\usepackage{amsfonts}
\usepackage{amsmath}
\usepackage{amssymb}
\usepackage{amsthm}
\usepackage{bm}
\usepackage[mathscr]{eucal}
\usepackage{bbold}
\usepackage{braket}
\usepackage{color}
\usepackage{comment}
\usepackage{dsfont}
\usepackage{jheppub}
\usepackage{framed}
\usepackage{physics}
\usepackage[normalem]{ulem}
\usepackage{tensor}
\usepackage{thmtools}
\usepackage{thm-restate}
\usepackage{enumitem}
\usepackage{pifont}
\usepackage{ulem}
\usepackage{tikz}
\usepackage{tikz-cd}
\usetikzlibrary{math} 
\usepackage{graphicx}
\usepackage{caption}
\usepackage{subcaption}
\usepackage{hyperref}
\def\la{\langle}
\def\ra{\rangle}
\def\tr{\operatorname{tr}}
\makeatletter\def\@fpheader{~}\makeatother
\allowdisplaybreaks  
\numberwithin{equation}{section}
\usepackage{titlesec}
\titleclass{\subsubsubsection}{straight}[\subsection]
\newcounter{subsubsubsection}[subsubsection]
\renewcommand\thesubsubsubsection{\thesubsubsection.\arabic{subsubsubsection}}
 
\titleformat{\subsubsubsection}
  {\normalfont\normalsize\bfseries}{\thesubsubsubsection}{1em}{}
\titlespacing*{\subsubsubsection}
{0pt}{3.25ex plus 1ex minus .2ex}{1.5ex plus .2ex}
\makeatletter
\renewcommand\paragraph{\@startsection{paragraph}{5}{\z@}%
  {3.25ex \@plus1ex \@minus.2ex}%
  {-1em}%
  {\normalfont\normalsize\bfseries}}
\renewcommand\subparagraph{\@startsection{subparagraph}{6}{\parindent}%
  {3.25ex \@plus1ex \@minus .2ex}%
  {-1em}%
  {\normalfont\normalsize\bfseries}}
\def\toclevel@subsubsubsection{4}
\def\toclevel@paragraph{5}
\def\toclevel@paragraph{6}
\def\l@subsubsubsection{\@dottedtocline{4}{7em}{4em}}
\def\l@paragraph{\@dottedtocline{5}{10em}{5em}}
\def\l@subparagraph{\@dottedtocline{6}{14em}{6em}}
\newtheorem{theorem}{Theorem}
\newtheorem{lemma}[theorem]{Lemma} 

\theoremstyle{definition}

\def\A{\mathcal{A}}
\def\H{\mathcal{H}}
\def\M{\mathcal{M}}
\def\N{\mathcal{N}}
\def\C{\mathcal{C}}
\def\la{\langle}
\def\ra{\rangle}
\def\barL{\bar L}
\def\barR{\bar R}
\def\tp{\otimes}
\def\til#1{\widetilde{#1}}

\def\b{\beta}
\def\g{\gamma}
\def\d{\delta}
\def\e{\epsilon}

\def\r{\rho}

\def\s{\sigma}

\def\t{\tau}

\def\ps{\psi}
\def\w{\omega}

\def\D{\Delta}

\def\P{\Pi}

\def\Ps{\Psi}

\title{Gravitational Algebras with Two Areas}
\author{Xuchen Cao, Thomas Faulkner, Zhencheng Wang}
\affiliation[]{Department of Physics, University of Illinois Urbana-Champaign, Urbana, IL 61801, USA}
\emailAdd{xuchenc2@illinois.edu}
\emailAdd{tomf@illinois.edu}
\emailAdd{zcwang1@illinois.edu}

\abstract{
We study gravitational algebras on spacetimes with two extremal surfaces. In the example of a long wormhole with two asymptotic AdS boundaries and two compact extremal surfaces, we discuss the assignment of gravitational algebras to various regions bounded by the extremal surfaces and/or asymptotic boundaries. 
Using the split property, we construct type II algebras from the crossed product in the left exterior, right exterior, the middle ``python's lunch'' region, 
and their complement regions. 
We also study the case where only the area sum operator or area difference operator is included as part of the gravitational algebra. This can be achieved by picking the appropriate microcanonical ensemble, and these gravitational algebras can either be type II or type III depending on the region. 
In the case where we include only the area difference mode, the crossed product gives rise to a weight that restricts to a trace on the middle region.
Differences of relative entropies with respect to this weight give differences in generalized entropies. This provides an algebraic understanding of the order parameter that controls the phase transitions between entanglement wedges.  We emphasize the role of operator-valued weights used in our construction.
}

\begin{document}

\maketitle

\section{Introduction}
\label{sec:intro}
Recently there has been significant progress in constructing gauge-invariant algebras in perturbative quantum gravity using a mathematical procedure called the modular crossed product \cite{Witten:2021unn,Chandrasekaran:2022cip, Chandrasekaran:2022eqq,Jensen:2023yxy,AliAhmad:2023etg,Kudler-Flam:2023qfl,Chen:2024rpx,Speranza:2025joj,Faulkner:2024gst,Kudler-Flam:2024psh,Klinger:2023auu,Klinger:2023tgi,AliAhmad:2024wja}. 
Physically, given the algebra of quantum field theory in some local region and an extra continuous degree of freedom coming either from a gravitational mode or from some external observer, the crossed product produces an algebra that satisfies the gravitational constraints. 

Such a crossed product construction has been shown to work in various cases, such as the exterior regions of black holes \cite{Witten:2021unn,Chandrasekaran:2022eqq,Kudler-Flam:2023qfl}, de Sitter space and other cosmological setups \cite{Chandrasekaran:2022cip,Kudler-Flam:2024psh,Chen:2024rpx}, general local spacetime regions \cite{Jensen:2023yxy,AliAhmad:2023etg,Chen:2024rpx}, etc. In the example of the exterior region of an AdS-Schwarzschild black hole, the crossed product includes an extra mode that represents the fluctuations of the right ADM Hamiltonian. This mode is formally a sum of the horizon area and the (one-sided) modular Hamiltonian of the Hartle-Hawking state:
\begin{equation}
    H_{R} = \frac{\bm A}{4G} + \frac{\bm h_R}{\beta}.
\end{equation}
The latter description then generalizes to any region bounded by an extremal surface, where the area operator of the extremal surface, added to a modular Hamiltonian in that region, forms a well-defined operator.  Inclusion of this mode gives the modular crossed product algebra \cite{Chen:2024rpx}.

Furthermore, the resulting algebra 
is type II, where, unlike the type III$_1$ algebra for local regions in quantum field theory (QFT), density matrices and their entropies can be rigorously defined. To leading order in perturbation theory, the crossed-product entropy for semiclassical states has been shown to agree with the generalized entropy \cite{Witten:2021unn,Chandrasekaran:2022cip,Chandrasekaran:2022eqq,Kudler-Flam:2023qfl}
\begin{equation}
    S_{\rm gen}=\frac{A}{4G} + S_{\rm bulk}.
\end{equation}
This concretely demonstrates the idea that the generalized entropy is a well-defined quantity in quantum gravity even though each individual term may not be \cite{Susskind:1994sm}\footnote{See \cite{Gesteau:2023hbq,Dong:2025crd} for some recent discussions on this.}.

In this paper, we consider some further examples of such gravitational algebras and their assignment to
subregions of spacetimes that contain more than one homologous extremal surface. Such spacetimes have been of great interest in recent discussions in holography: they exhibit phase transitions in the holographic entanglement entropy and entanglement wedges \cite{Akers:2020pmf,Marolf:2020vsi,Dong:2020iod,Penington:2019npb,Almheiri:2019psf,Penington:2019kki,Almheiri:2019qdq}; the ``python's lunch'' region between two (locally minimal) extremal surfaces is important in the discussion of bulk-reconstruction complexity \cite{Brown:2019rox,Engelhardt:2021qjs,Engelhardt:2021mue}. This work will be a first step towards using an algebraic approach to study the above questions. 

\begin{figure}[h!]
  \begin{minipage}[c]{0.5\textwidth}
\centering
    \includegraphics[width=0.8\textwidth]{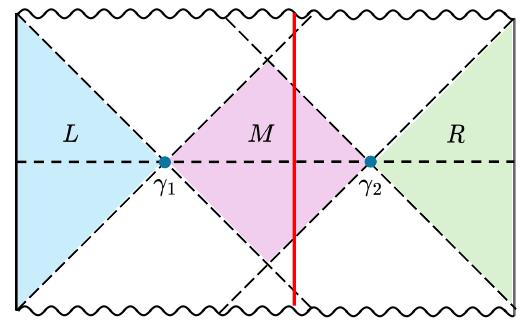}
\end{minipage}
  \begin{minipage}[c]{0.5\textwidth}
\centering
    \includegraphics[width=0.8\textwidth]{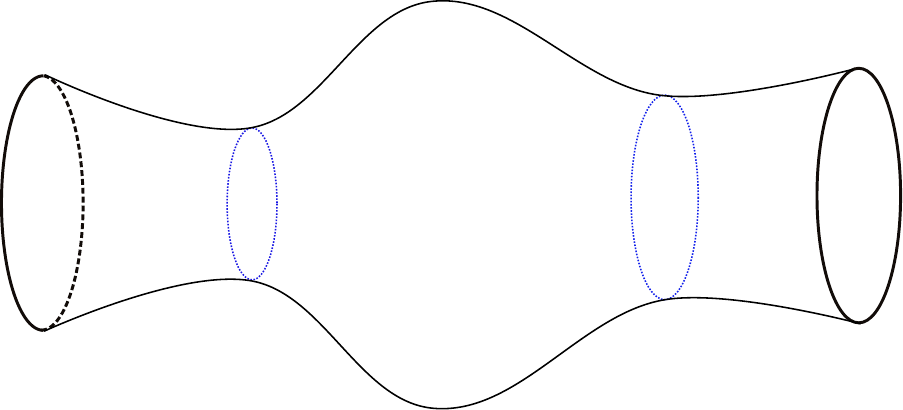}
\end{minipage}
  \caption{\textbf{Left:} The Penrose diagram of a long wormhole supported by a thin shell (the red vertical line). Geometries in the left and right exteriors (denoted as $L$ and $R$) are described by the Schwarzschild metric. There are two extremal surfaces (blue dots, denoted as $\gamma_1$ and $\gamma_2$), and the region they bound (denoted as $M$) is referred to as the python's lunch. The time-reflection-symmetric Cauchy slice in this geometry is denoted as the horizontal black dashed line. \textbf{Right:} The time-reflection-symmetric Cauchy slice of the long wormhole spacetime, where the two blue circles are the two extremal surfaces.}  
  \label{fig:PL-Penrose}
\end{figure}

For simplicity and concreteness, we will mainly focus on spacetimes that contain long wormholes, where there are two asymptotically AdS boundaries and two bifurcate horizons; see Figure \ref{fig:PL-Penrose}. As a further simplification, we will take the two exterior regions to be diffeomorphic to the exterior regions of Schwarzschild black holes with possibly different temperatures. One specific example of this is the long wormhole prepared by the Euclidean gravitational path integral where heavy operators inserted on the boundary excite thin shells behind the horizons in the bulk \cite{Goel:2018ubv,Balasubramanian:2022gmo}.

In such a long wormhole, the two extremal surfaces naturally divide any Cauchy slice that passes through them into three: left exterior, right exterior, and the middle (the python's lunch \cite{Brown:2019rox}). We are interested in the gravitational algebras for (the causal development of) these three regions and their complements. The relevant gravitational modes are the fluctuations of the two extremal-surface areas, as well as their canonical conjugates.

To rigorously define the gravitational algebras, we will work with the microcanonical ensembles, and also take the strict $G_N\to 0$ limit \cite{Chandrasekaran:2022eqq}\footnote{One can work with the canonical ensemble and consider a perturbative expansion in $G_N$ or $1/N$ \cite{Witten:2021unn}, however, technically one needs to consider algebras over the ring $\mathbb{C}[[1/N]]$ of formal power series rather than complex numbers.\label{fn2}}. In a microcanonical ensemble where the corresponding ADM Hamiltonian $H$ has $O(1)$ fluctuations, $X\equiv \beta(H-\la H\ra)$ is a well-defined operator. For a long wormhole, there are two independent ADM Hamiltonians $H_L$ and $H_R$. Formally, they correspond to the area of the left and right extremal surface $A_1$ and $A_2$ respectively. 

We first discuss the gravitational algebras when both areas are included. This can be realized by the microcanonical ensemble for both $H_L$ and $H_R$. Here the gravitational Hilbert space is $\H_{QFT} \otimes L^2(\mathbb{R}_{X_1}) \otimes L^2(\mathbb{R}_{X_2})$, where $X_1 \equiv \beta_1(H_L-\la H_L \ra)$ and $X_2 \equiv \beta_2(H_R-\la H_R \ra)$. To construct the gravitational algebras, we make use of the split property in quantum field theory \cite{DLsplit84}. Using the split property, we can find an isomorphism that maps the bulk QFT Hilbert space on the long wormhole to the tensor product of bulk QFT Hilbert spaces on a pair of two-sided Schwarzschild black holes. The local QFT algebras on the exterior of the wormhole are also related, using this isomorphism, to local algebras on the pair of two-sided black holes. 


In the split picture, it is straightforward to construct type II$_\infty$ crossed-product algebras in the exterior regions, closely following \cite{Witten:2021unn,Chandrasekaran:2022eqq}. We are also able to construct the crossed-product algebra in the python's lunch. This is done by doing two crossed products in the split picture, including two modes, each of which is a sum of the extremal-surface area and an appropriate modular Hamiltonian\footnote{A similar use of the split property was presented in \cite{Chen:2024rpx} in the construction of crossed-product algebras in the exterior region of Schwarzschild-de Sitter black holes. The difference between our construction and the one in \cite{Chen:2024rpx} is that, there are two independent gravitational degrees of freedom that are relevant, instead of one in their case.}. For a semiclassical state, its entropy on the crossed-product algebras matches the generalized entropy in each region.  

We then move on to consider the gravitational algebras where only the area sum $A_1+A_2$ or the area difference $A_1-A_2$ is the dynamical gravitational degree of freedom. This can be realized in the microcanonical ensemble for only $H_L+H_R$ or $H_L-H_R$. For this discussion, it is more convenient to reorganize the gravitational Hilbert space to be $\H_{QFT}\otimes L^2(\mathbb{R}_{X_+})\otimes L^2(\mathbb{R}_{X_-})$, where $X_+\equiv X_1+X_2$ and $X_-\equiv X_1-X_2$.

For the $H_L+H_R$ microcanonical ensemble, while the fluctuation of $X_+$ is $O(1)$, $X_-/N$ becomes central for the gravitational algebras. Since this center exists in all algebras, and it only acts on the tensor factor $L^2(\mathbb{R}_{X_-})$ in the Hilbert space, the non-trivial part of the gravitational algebras only involves the quantum fields and $X_+$. Factoring out the common mode $X_-/N$, we find that the gravitational algebra is type II$_\infty$ in the python's lunch, but type III$_1$ for the left or the right exterior region. Physically, this is because the python's lunch region can access both areas, and thus the area sum; while the left or the right exterior regions alone can only access one of the areas, but not the sum.

For the $H_L-H_R$ microcanonical ensemble, we can similarly factor out the $X_+/N$ mode, and discuss the gravitational algebras involving only QFT degrees of freedom and $X_-$. To construct the gravitational algebras in this case, the main difficulty is to find a desired modular flow compatible with the area difference. Take for example the python's lunch region, a generic modular flow will resemble an upward boost near both edges and therefore cannot combine with $A_1-A_2$ to form a well-defined operator. 

However, an appropriate modular flow can be found using operator-valued weights \cite{HAAGERUP1979a,HAAGERUP1979b}. For an inclusion of algebras, the operator-valued weight is a generalization of the conditional expectation in a sense that it is not normalizable. When there exists an operator-valued weight $T:\M \to \N$ for $\N \subset \M$, for any state $\omega$ on $\N$, we can construct a weight $\omega \circ T$ on $\M$, and this weight has the property that its modular flow on $\M$ will keep $\N$ fixed and the relative commutant $\N' \wedge \M$ fixed (as sets). Using the split property, it is well known that an operator-valued weight exists for the algebra inclusion $\A_R \subset \A_{M\cup R}=\A_M\vee \A_R$ (and similarly $\A_L \subset \A_{L\cup M}=\A_L \vee \A_M$), where $M$ refers to the middle python's lunch region. Then from the property of operator-valued weights, we find a modular flow that is an upward boost near one edge, and a downward boost near the other edge. So together with $A_1-A_2$, it produces a well-defined operator, and can be used to construct a type II crossed product algebra. 

The structure of gravitational algebras in the $H_L-H_R$ microcanonical ensemble is similar to the $H_L+H_R$ microcanonical ensemble, except that the modular flow used here is constructed from an operator-valued weight. In this case, the physical interpretation of the type II entropy in the python's lunch remains obscure. However, another quantity is of physical importance: the difference in generalized entropies for region $MR\equiv M\cup R$ and $R$. This is important since it is the order parameter for the phase transition in the entanglement wedge. 

Given a semiclassical state $\widetilde{\omega}_{\Xi}$, a linear functional on the gravitational algebra for $MR$,  we show that the generalized entropy difference, up to an additive constant, can actually be computed from the difference of relative entropies for $\widetilde{\omega}_{\Xi}$
\begin{equation}
\label{sdiff}
     S_{\rm rel}(\widetilde{\omega}_{\Xi}|\widetilde{\omega}_{2} \circ \widetilde{T})_{MR}-S_{\rm rel}(\widetilde{\omega}_{\Xi}|\widetilde{\omega}_{2})_{R}
    = S_{\rm rel}(\widetilde{\omega}_{\Xi}| \widetilde{\omega}_{\Xi} \circ \widetilde{T})_{MR}
\end{equation} 
where both terms on the left hand side are defined on the corresponding type III$_1$ gravitational algebra for $MR$ and $R$. Here $\widetilde{\omega}_{2}$ is any state on $R$, and $\widetilde{T}$ is the gravitational version of the
operator-valued weight $T$ from $MR$ to $R$. $\widetilde{T}$
has the special property that 
$\widetilde{\omega}_{2} \circ \widetilde{T}$ reduces to a maximally mixed (i.e. tracial) weight on the type II$_\infty$ gravitational algebra of $M$. The second expression in \eqref{sdiff} is the analog of the Petz formula for conditional expectations \cite{ohya2004quantum}. To our knowledge this has not generally been proven, but we provide a proof for the special case studied here. Hence the generalized entropy difference reduces to a somewhat familiar expression for conditional entropy \cite{gao2020relative} which also appears in the entropic certainty relations of \cite{Casini:2019kex,Magan:2020ake}. We also expect this calculation to be generalized to spacetimes with boundary-anchored extremal surfaces, where only the area difference is a well-defined operator. 

The paper is organized as follows. We start by reviewing the crossed product construction and its application in gravity in Section~\ref{sec:review}, focusing on the example of the exterior regions of Schwarzschild black holes. Some technical details are deferred to Appendices \ref{app:a} and \ref{app:b}. In Section~\ref{sec:ovw}, we review the mathematical notion of operator-valued weight and split property in quantum field theory, which will be useful for later constructions. We also prove the existence of operator-valued weights from split property for certain nested regions in quantum field theory. Sections~\ref{sec:pythonslunch} and \ref{sec:singlearea} are devoted to constructions of crossed product algebras for different regions in the long wormhole spacetime, as well as discussions on entropies in those type II$_\infty$ algebras. We first do the construction in the case where both area modes are accessible, and in the case where only the sum mode is accessible. Then we discuss the crossed product algebra with only the difference mode, making use of the operator-valued weight. Finally, we conclude the paper with discussions in Section~\ref{sec:discussion}, including some speculations on the relation between operator-valued weights and the decoding complexity and non-isometric code in python's lunch.

\paragraph{Notations and Assumptions:} 
In this paper, for simplicity, we will use the same notation for a region on a Cauchy slice and its domain of dependence. For example, for a region $R$ on a Cauchy slice, its domain of dependence will also be labeled $R$. If there are two regions $M$ and $R$, we will also use a shorthand notation $MR=M\cup R $ to denote both the region that is the union of $M$ and $R$ on a Cauchy slice and the domain of dependence of $M\cup R$ (which is generally larger than the union of the two individual domains of dependence). 

When discussing algebras for a region $R$, we will use $\A_R$ for its quantum field theory algebra on this background, and use $\widetilde \A_R$ for its gravitational algebra when certain gravitational degrees of freedom are included. We also assume the QFT algebras satisfy all the nice properties of a complete theory, such as additivity and Haag duality. For example, these can be used to derive the relationship: 
\begin{equation}
(\mathcal{A}_M \vee \mathcal{A}_R) \wedge \mathcal{A}_R' = \mathcal{A}_M.
\end{equation}

In this paper, depending on the context, a state can either refer to a vector in the Hilbert space or a normal linear functional on the appropriate algebra. The qualifier normal, faithful and semifinite in the case of states, weights, conditional expectations and operator-valued weights will be assumed without mention unless stated otherwise. For a modular operator $\Delta_{\psi;\A}$, it is labeled by a vector $|\psi\rangle$ and an algebra $\A$. For a relative modular operator $\Delta_{\psi,\phi;\A}$, it is labeled by two vectors $|\psi\rangle,|\phi\rangle$ and an algebra $\A$. We will omit the $\A$ label when it is clear from the context what the relevant algebra is. The relative modular operators end up only depending on the corresponding induced states as linear functionals on the algebra and its commutant. As such, sometimes we will label the relative modular operators with only the corresponding induced states. Lastly, we will use boldface letters like $\bm h$ for quantities that are not well-defined in the quantum theory, but are helpful for physical intuitions.

\section{Review of crossed products in gravity}
\label{sec:review}

In this section, we review the mathematical notion of the crossed product and its application in perturbative quantum gravity. Aside from referencing our specific notation, those with knowledge of this subject are encouraged to skip ahead to Section~\ref{sec:ovw}.
As an example, we specifically review the crossed-product algebras for exterior regions of two-sided AdS-Schwarzschild black holes \cite{Witten:2021unn,Chandrasekaran:2022eqq} in Section \ref{sec:2.1}, and discuss how the generalized entropy can be derived from the entropy of the resulting type II algebra in Section \ref{subsec:BHentropy}. We then discuss the Hilbert spaces and algebras for a general region bounded by an extremal surface in a general spacetime in Section \ref{subsec:generalextremal}. 

\subsection{Gravitational dressing and crossed products}
\label{sec:2.1}

Consider a two-sided AdS-Schwarzschild black hole with temperature $T=1/\beta$ as a solution to the Einstein equations. This solution is non-unique due to the existence of the time-shift mode (see e.g. \cite{Chandrasekaran:2022eqq,Harlow:2018tqv}). This time shift is defined in the following way: consider a timeslice $\C$ with fixed Schwarzschild time, extending from the right boundary time $t_R$ to the bifurcation horizon. $\C$ has zero extrinsic curvature and intersects the right boundary perpendicularly, which can then be extended into a unique zero-extrinsic-curvature surface which continues to the left boundary and intersects the left boundary at boundary time $t_L$. The time shift is then defined to be $t=t_L+t_R$\footnote{Here we define boundary times to flow in the same direction on both boundaries.}.

\begin{figure}[h!]
  \centering
  \includegraphics[width=0.4\textwidth]{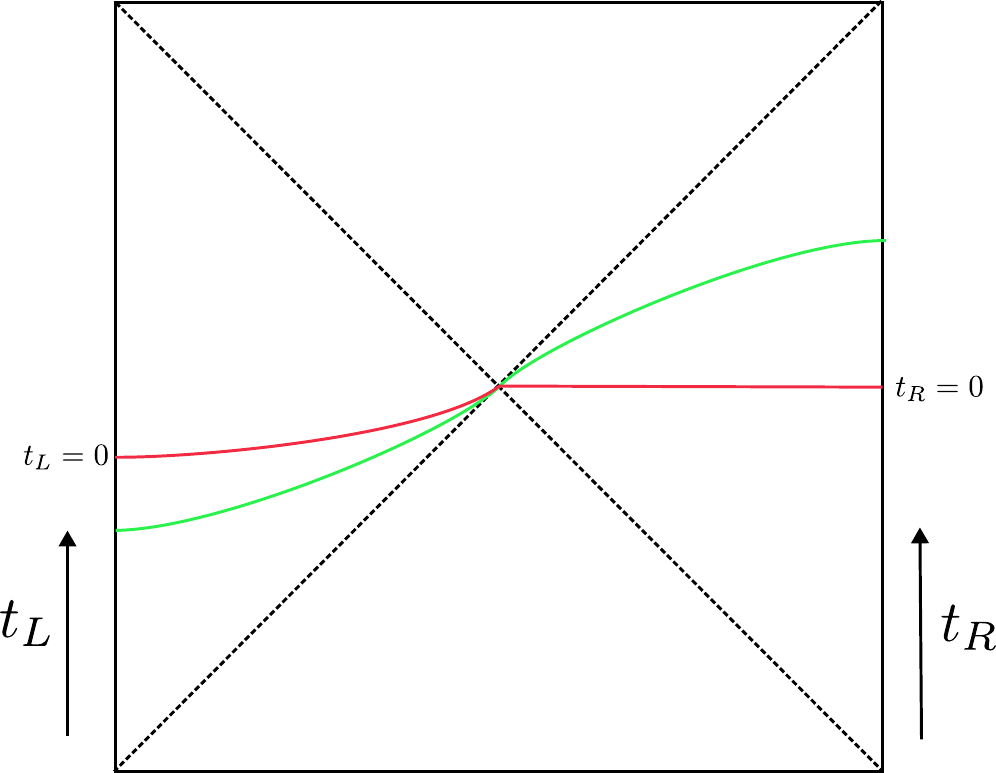} 
  \caption{Timeslices in two-sided AdS-Schwarzschild black hole with a non-zero time shift. The green surface denotes the zero-extrinsic-curvature timeslice which is smooth at the bifurcation horizon. The red surface is a timeslice intersecting boundaries at $t_L=t_R=0$, which has a kink at the horizon. }
  \label{timeslice} 
\end{figure}

The two-sided boost symmetry of the Schwarzschild solution ensures that $t$ is independent of the choice of $t_R$. Another way to understand this time shift is to consider a timeslice $\C_0$ with constant Schwarzschild times in both left and right exteriors and intersects the two boundaries at $t_L=t_R=0$. For a general solution with $t\neq 0$, $\C_0$ necessarily has a kink at the bifurcation horizon, where the extrinsic curvature diverges. That is, we cannot have zero extrinsic curvature everywhere if we impose the condition $t_L=t_R=0$. See Figure~\ref{timeslice} for an illustration.
On the gravitational phase space, the time shift $t$ is conjugate to left ADM Hamiltonian $H_{L}$, or the right ADM Hamiltonian $H_R$.  This is because $H_R$ and $H_L$ are subject to the constraint $H_R-H_L=\hat h/\beta$ where $\hat h$ is the two-sided boost generator in the bulk. Due to this constraint only one of the $H_L$ and $H_R$ is independent and we thus have only one pair of conjugate variables. Furthermore, $t_L-t_R$ is not a dynamical variable as a result of this constraint. Solutions with different $t_L-t_R$ are gauge equivalent with $H_R-H_L-\hat h/\beta$ the generator of gauge transformations. By the uncertainty relation they satisfy $\delta t\delta H\sim \mathcal O(1)$.

There are two situations we can consider for perturbative gravitational fluctuations of the spacetime. In the canonical ensemble, $t$ has $O(\sqrt{G_N})$ fluctuations and ADM Hamiltonians have  $O(\frac{1}{\sqrt{G_N}})$ fluctuations; while in the microcanonical ensemble, fluctuations of the ADM Hamiltonian are $O(1)$, thus the fluctuation of $t$ is also of order $O(1)$. In Appendix~\ref{app:a} we explain that this difference is a result of considering different dual states in the boundary CFT. As mentioned in the Introduction, we will work with the microcanonical ensemble for both simplicity and mathematical rigor (see Footnote \ref{fn2}).

Let $\A_{L}$ and $\A_{R}$ be the QFT algebras in left and right exteriors, which are both type III$_1$ von Neumann algebras, with $\A_L'=\A_R$. We denote as $\H_{QFT}$ the Hilbert space of the quantum field theory on which $\A_{R}$ and $\A_{L}$ act. For gravitational algebras, one should incorporate spacetime fluctuations, including perturbative local graviton modes which can be dealt with using standard gauge fixing procedures. To take into account fluctuations of the ADM Hamiltonian which is a global gravitational mode, we need to choose how to specify QFT states for each time shift $t$. The gravitational Hilbert space is a direct sum over the QFT Hilbert spaces
for each $t \in \mathbb{R}$. We can then identify the different Hilbert spaces by identifying their states. One possible choice is to first pick a zero-extrinsic-curvature timeslice with fixed $t_R$ as discussed above, for example we can choose $t_R=0$. At the classical phase space level, we can specify a state by fixing its field configuration on this specific timeslice\footnote{Strictly speaking, we have to define the state in an infinitesimal neighborhood of the slice. Global hyperbolicity ensures that these data are enough to specify a state.}. In quantum theory, we specify a state by fixing all correlation functions with reference to this timeslice. That is, when we write down correlators such as $\expval{\phi(x_1,t_1)\phi(x_2,t_2)\ldots}$, the bulk coordinates $(t_i,x_i)$ are measured in relation to this timeslice. 

 This procedure is usually referred to as gravitational dressing to the right boundary. Now we have an unambiguous action of $a_R \in \A_{R}$ on $\H = \oplus_t \mathcal{H}_{QFT} \cong \H_{QFT} \otimes L^2(\mathbb{R})$ given by
$a_R \otimes 1$ since these operators must commute with the time shift mode, represented here as the momentum operator $\Pi$ on $L^2(\mathbb{R})$.

In addition to $\A_{R} \otimes 1$, we also need to add the fluctuation of the right ADM Hamiltonian $H_R'=H_R-\langle H_R\rangle$ as an observable. 
Note that $H_R=H_L+\hat h/\beta$, where $H_L$ acts on the left boundary and thus commutes with $\A_{R} \otimes 1$ implying
that we can identify $\beta(H_L-\langle H_L \rangle)$
with $X$, the canonical conjugate of $\Pi$, on $L^2(\mathbb{R})$. The algebra is hence\footnote{We denote by $'$ the commutant of an algebra, which is defined to be the set of all bounded operators on $\cal H$ which commute with all elements in $\mathcal A$. Here we take the double commutant of a $\star$-set of operators, which is equivalent to taking the completion of the algebra generated by this $\star$-set, and with respect to weak operator topology.}
\begin{equation}
    \widetilde\A_R=\{\A_{R},X+\hat h\}''.
\end{equation}
Recall that $\hat h$ generates an automorphism on $\A_{R}$ as it is the boost generator. In this case $\widetilde \A_R$ is called a \emph{crossed product} \cite{takesaki2003theory}. Furthermore, it is well known that $\hat h$ can be written as the modular Hamiltonian of the Hartle-Hawking state. Thus $A_R$ is a \emph{modular crossed product}, denoted as $\widetilde\A_R=\A_{R}\rtimes_{\s}\mathbb R$, which turns out to be a type II$_\infty$ algebra.

Next we discuss the algebra defined in the left exterior. Since we have made the choice to dress to the right boundary, the definition of left operators becomes more complicated. For example, if we let $\A_{L}$ act on $\H_{QFT}$ directly, then it does not commute with $X+\hat h\in\widetilde\A_{R}$, which contradicts the requirement that left and right operators should commute as they are causally disconnected. The root of this contradiction is that when we act $\A_{L}$ directly on $\H_{QFT}$, we are in fact dressing left variables to the right boundary, and thus these dressed operators are not purely defined in the left. The resolution is to switch the dressing to the left boundary, that is to define states with respect to the zero-extrinsic-curvature timeslice with $t_L=0$. The new dressing leads to another factorization of the overall Hilbert space $\H=\H_{QFT}\otimes L^2(\mathbb R)$, where the $L^2(\mathbb R)$ factor is now acted on by $H_R$ instead of $H_L$. The two different dressing schemes are related by a unitary transformation $e^{-i\hat ht} \equiv
e^{-i\hat h\Pi}$.
To see this note that the unitary boosts the state on the $t_R=0$ timeslice to the one on the $t_L=0$ slice, and exchanges $H_L$ and $H_R$ as $t$ is conjugate to both of them. Therefore, if we choose to dress to the right boundary, the correct way to write the left algebra is 
\begin{equation}
    \widetilde\A_L=\{e^{i\Pi\hat h}\mathcal A_{L}e^{-i\Pi \hat h},X\}''
\end{equation}
where we have rewritten $t$ as $\P$ to emphasize that it is the conjugate momentum of $X$. It can be easily verified that $\widetilde\A_L$ and $\widetilde\A_R$ are indeed commutants of each other. We emphasize that the asymmetry in the forms of $\widetilde\A_L$ and $\widetilde\A_R$ is the result of the specific gravitational dressing we chose. In principle we can choose some specific dressing to make them symmetric, but the physical significance of such a dressing is not so clear.

There is yet another useful perspective to view this construction which we will refer to frequently in the remainder of this paper \cite{Chen:2024rpx}. We exploit the fact that the left and right ADM Hamiltonians can be formally written as 
\begin{equation}\label{formal decomposition}
     H_{L,R}=\frac{\bm{\d A}}{4G}+\frac{\bm h_{L,R}}{\beta}.
\end{equation}
Here $\bm{\d A}$ is the area fluctuation and $\bm h_{L,R}$ are the one-sided boost generators in the left(right) exteriors. Since $\hat h=\bm h_R-\bm h_L$, the constraint $H_R-H_L=\hat h/\beta$ is automatically satisfied. We call this a formal decomposition as both $\frac{\bm {\d A}}{4G}$ and $\bm h_{L,R}$ are in fact not operators as they have divergent fluctuations. In the following we will use boldface letters to emphasize quantities which are only formally written as operators. It can be shown that the HRT area operator $\frac{\bm {\d A}}{4G}$ is the one which creates kinks in a given Cauchy slice~\cite{Bousso:2020yxi,Kaplan:2022orm}, which are exactly the time shifts in our case. The advantage of this decomposition is that it makes clear to which region an observable belongs to. For example, $\bm{\d A}$ is accessible to both left and right exterior regions as it is the area of the shared boundary between left and right. On the other hand, $\bm h_L$ and $\bm h_R$ are defined in the left and right exteriors, so we conclude that $H_L$ and $H_R$ belong to the left and right exteriors respectively. Another important observation is that the singularity of the one-sided boosts $\bm h_{L,R}$ comes from the near-horizon region where they are discontinuous. This singularity exactly cancels the singularity of $\frac{\bm {\d A}}{4G}$. Note that this cancellation of divergence only depends on the singularities of $\bm h_{L,R}$ at the horizon, we can therefore replace $\bm h_{L,R}$ by any other operator with the same singularity to define a good operator. 

It has been conjectured that under some reasonable conditions the one-sided modular flow of an arbitrary state always approaches a boost in the vicinity of the horizon, as any QFT state should look like vacuum at sufficiently short distances. This can be justified by the following mathematical fact, let $\mathcal A$ be a type $\text{III}_1$ algebra associated to a subregion in the spacetime and $\psi,\phi$ be two different normal, semifinite and faithful states on $\A$. It can be proved that the following Connes cocycle operator  
\begin{equation}
    u_{\varphi,\psi}(t)=\Delta_{\phi,\psi}^{it}\Delta_\psi^{-it}
\end{equation}
is an element of the algebra $\mathcal A$~\cite{takesaki2003theory}. Here $\D_{\phi,\psi}$ is the relative modular operator between $\phi$ and $\psi$\footnote{For two arbitrary purified states $\ket{\ps}$ and $\ket{\eta}$ on a von Neumann algebra $\A$, we can define the antilinear relative Tomita operator $S_{\ps,\eta}$ by $S_{\ps,\eta}a\ket{\eta}=a^\dagger \ket{\ps}$, $\forall a\in\A$. The relative modular operator is then defined by $\D_{\ps,\eta}=S^\dagger_{\ps,\eta}S_{\ps,\eta}$. A notation that is unambiguous would label the algebra $\D_{\psi,\eta;\A}$, but for simplicity, and when the algebra is understood, we will often drop this label.  Another possible unambiguous notation, which we will use later in Section~\ref{sec:5.2}, is to label the relative modular operators by the linear functional states, one for the algebra and the commutant $\Delta_{\rho,\sigma}$ where $\rho = \left< \psi \right|( \cdot)|_{\mathcal{A}} \left| \psi \right>$ and $\sigma = \left< \eta \right|(\cdot)|_{\mathcal{A}'}\left| \eta \right>$. 
}.  and $\D_\psi$ is the modular operator for the state $\psi$. Formally the cocycle can be written as
\begin{equation}
    u_{\varphi,\psi}(t)=(\bm\rho_{\varphi}^A)^{it}(\bm\rho_{\psi}^A)^{-it}\otimes \mathbb I_{\bar A}
\end{equation}
where $\bm\rho_\psi^A$ is the reduced density matrix of state $\psi$ on $A$ while $\bm h_{\varphi}^A$ is the corresponding one-sided modular Hamiltonian on $A$. This tells us that for two different states, despite the fact that their one-sided modular Hamiltonians are both singular and therefore not well defined operators, their difference is still a well-behaved operator as the singularities at the horizon cancel.

\subsection{Generalized entropy is von Neumann entropy}
\label{subsec:BHentropy}
For the left and right exteriors, the crossed product construction gives the type $\text{II}_\infty$ algebras $\widetilde \A_L$ and $\widetilde \A_R$. Each algebra has a trace, and consequently density matrices and entropies. We will calculate the entropy in these type $\text{II}_\infty$ algebras for semiclassical states, and show that they agree with the generalized entropy. The proof exploits the formal decomposition~\eqref{formal decomposition} above. The same strategy will be used in the calculation of entropies for the type $\text{II}_\infty$ algebra in the long wormhole later. Here we show the calculation for $\widetilde\A_R$, and the entropy calculation for $\widetilde\A_L$ can be shown similarly, despite some slight differences due to the asymmetry in the forms of $\widetilde\A_L$ and $\widetilde\A_R$.

First we need the definition of the trace. We will use $\ket{\eta}$ to denote the state which we use to construct the modular crossed product, which we take to be the Hartle-Hawking state. The trace can be formally written as expectation value in the following non-normalizable state (See Appendix~\ref{app:b} for details)  
\begin{equation}\label{trace}
    \ket{\t}=\int dX\;e^{\frac{X}{2}}\ket{\eta}\otimes\ket{X}.
\end{equation}
We should emphasize here that due to the divergent factor $e^{\frac{X}{2}}$ in the definition of the trace, some operators (for example, the identity operator) have divergent trace. Technically, $\ket{\t}$ defines a tracial weight instead of a state. 

Next we discuss the expression of the density matrix for semiclassical states of the following form (sometimes referred to as classical-quantum states)
\begin{equation}
    \ket{\Ps}=\int dX \; g(X)\ket{\ps}\otimes\ket{X},
\end{equation}
where $g(X)$ is normalized as $\int dX |g(X)|^2=1$, and it is a slowly varying function $g(X)= \epsilon^{\frac{1}{2}} \mathcal{G} (\epsilon X)$ with $\epsilon \ll 1$. Therefore, the time shift of this state $t \sim O(\epsilon)$. 

It is easy to check that the following expression is the approximate density operator up to order $\e$ corrections (See Appendix~\ref{app:b})
\begin{equation}\label{density}
    \r_\Ps=|g(X)|^2e^{-X}\Delta_{\ps,\eta}
\end{equation}
by verifying the condition $\tr (\rho_\Psi \hat a)=\expval{\rho_{\Ps}\hat a}{\t}=\expval{\hat a}{\Ps}$ up to $O(\epsilon)$ corrections, where $\Delta_{\ps,\eta}$ is the relative modular operator between states $\ket{\ps}$ and $\ket{\eta}$.

With the density matrix at hand we are now ready to calculate the entropy, the type $\text{II}$ entropy is 
\begin{align}
    S&=-\expval*{\ln \rho_\Ps}{\Ps}\nonumber\\
    &=-\int dX |g(X)|^2\ln|g(X)|^2+\expval*{X}{\Ps}-\expval{\ln \Delta_{\ps,\eta}}{\ps}.
\end{align}
To see that the above entropy formula actually corresponds to the generalized entropy, we implement the formal decomposition~\eqref{formal decomposition}. Recall that we are dealing with the right exterior of the black hole, and $X$ should be understood as the ADM Hamiltonian fluctuation on the left boundary, so the second term can be formally written as
\begin{equation}
    \expval*{X}{\Ps}=\expval{\frac{\bm{\d A}}{4G}}_\Ps+\expval{\bm h_L}{\ps}.
\end{equation}
On the other hand, we can formally write 
\begin{equation}
    \ln \Delta_{\ps,\eta}=\ln\big[\bm\rho_{\ps}^R\otimes(\bm\rho_{\eta}^L)^{-1}\big]=\ln \bm\rho_{\ps}^R-\ln\bm\rho_{\eta}^L=\ln \bm\rho_{\ps}^R+\bm h_L,
\end{equation}
where superscripts $L,R$ denote that these are density matrices in the left and right exterior. Note that for the Hartle-Hawking state $\ket{\eta}$ the density operator is $\bm\rho_{\eta}^L=e^{-\bm h_L}$, where $\bm h_L$ is the left boost generator. Combining the terms above we have
\begin{equation}
    S=\expval{\frac{\bm {\d A}}{4G}}_\Ps-\expval{\ln \bm \rho_{\ps}^R}{\ps}-\int dX |g(X)|^2\ln|g(X)|^2.
\end{equation}
In the above expression, the first term is the expectation value of the horizon area, the second is the QFT entropy for the right exterior, and the third term comes from area fluctuations. Thus the expression agrees with the generalized entropy \cite{Bekenstein:1973ur}. Here we see that the decomposition~\eqref{formal decomposition} drastically simplifies the proof, which was originally done by invoking the Raychaudhuri equation in~\cite{Chandrasekaran:2022eqq}. This trick will be used repeatedly to calculate the type II entropy for the python's lunch in the long wormhole.

\subsection{Crossed product in general spacetimes with extremal surfaces}
\label{subsec:generalextremal}

Next we discuss the crossed product algebra for a region $S$ bounded by an extremal surface $\gamma$ in general spacetimes. We follow and expand on the discussion in \cite{Chen:2024rpx}.

We first discuss the classical phase space associated to an extremal surface. In Einstein gravity, the phase space for a given Cauchy slice is spanned by the induced metric $h_{ij}$, and the conjugate momenta $\Pi_{ij}$ (or equivalently the extrinsic curvature $K_{ij}$). Given an extremal surface $\gamma$, since its area $A_\gamma$ is part of $h_{ij}$, it has vanishing Poisson brackets with $h_{ij}$, so we only need to look at how the action of $A_\gamma$ affects $K_{ij}$. It has long been argued that the classical action of the area of any codimension-2 surface generates a ``kink'' (relative boost between the two sides of $\gamma$) on it \cite{Carlip:1993sa,Donnelly:2016auv,Chandrasekaran:2019ewn,Bousso:2020yxi}. For a general codimension-2 surface, the Poisson bracket between its area and $K_{ij}$ has been explicitly worked out in \cite{Kaplan:2022orm}, 
\begin{equation}
    \{\frac{A_\gamma}{4G}, K_{ij} \} = -2 \pi \hat{\delta}_{\Sigma}(\gamma, x) \perp^i \perp^j .
\end{equation}
where $\perp^i$ is the unit normal to $\gamma$ in $\Sigma$ and $\hat \delta_\Sigma(\gamma,x)$ is a one-dimensional delta function of the proper distance between $x$ and $\gamma$ measured along geodesics in $\Sigma$ orthogonal to $\gamma$. 

For an extremal surface $\gamma$, the initial data after the action of the area still satisfy the gravitational constraints. But this is generally not true for non-extremal surfaces \cite{Bousso:2019dxk}. After the action of $e^{i A_\gamma T/4G}$, we obtain a Cauchy slice denoted by $\Sigma_T$\footnote{The discussion here generalizes to the case with multiple extremal surfaces, where we should have label $T_1,T_2\ldots$ instead of a single $T$. Here we use a single $T$ for simplicity.}. Since the data on $\Sigma_T$ satisfy the gravitational constraints, one can evolve $\Sigma_T$ using the Einstein equations in the presence of appropriate boundary conditions\footnote{This is especially important for boundary-anchored extremal surfaces.  Although the action of area generates a kink, it does not change the asymptotic AdS boundary condition \cite{Kaplan:2022orm,Bousso:2020yxi}.} to obtain the full spacetime solution, which we call $M_T$. Since the action of $A_\gamma$ is localized on $\gamma$, the resulting spacetime $M_T$ will generally differ from the original spacetime in the future and past of $\gamma$, but stay the same in the domains of dependence of $S$ and $\bar S$. Note that $M_T$ is smooth in simple cases, but can have Weyl tensor shocks on the lightcones of the extremal surface $\gamma$ when $\gamma$ has a non-vanishing shear \cite{Bousso:2020yxi}.

However, there is a very special case when $M_T$ is the same as the original spacetime $M_0$, that is, when there exists a (generally non-smooth) Cauchy slice $\widetilde \Sigma_T$  in $M$, whose Cauchy data is diffeomorphic to the Cauchy data on $\Sigma_T$. In fact, as we will see, this is the case for the long wormhole we study in this paper.

\begin{figure}[h!]
  \centering
  \includegraphics[width=0.5\textwidth]{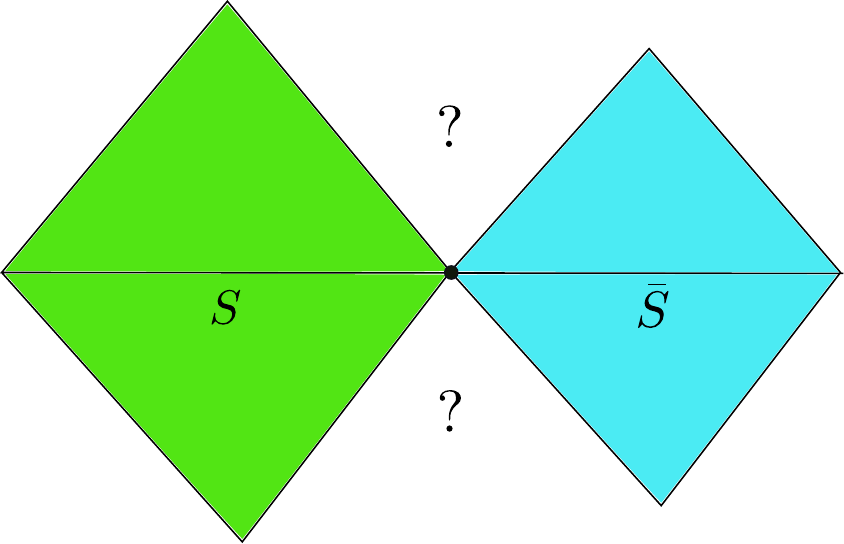} 
  \caption{Spacetime viewed as the development of Cauchy slices with different kink angles at the extremal surface (the black dot). Causality ensures that the domains of dependence of $A$ and $\bar A$ are independent of $t$, while the geometry outside depends on $t$ in general.}
  \label{time dev kink} 
\end{figure}

Next we discuss quantum field theories on spacetimes $M_T$. Defining a quantum field theory Hilbert space when $M_T$ has shockwaves is subtle, since picking out a natural class of states often requires a smoothness assumption on the underlying spacetime. This difficulty can presumably be overcome, and
we will discuss this issue more in Section \ref{sec:discussion}, deferring an in-depth study of this to future work. For the current discussion we only consider smooth $M_T$. For each $M_T$, there is a quantum field theory Hilbert space $\H_T$ defined on it. The overall Hilbert space is then a direct integral        
\begin{equation}
    \H=\int^{\oplus} dT\;\H_T.
\end{equation}

Now we want to find a natural way to identify the Hilbert spaces for different $\H_T$ so that we can write $\H=\H_0\tp L^2(\mathbb R)$. In quantum theory we expect $\mathcal{H}_T$ to furnish normal and standard representations  of $\A_S$ and $\A_{\bar S} = A_S'$ (defined as the von Neumann algebras associated to these regions on $M_0$.) It follows that there exists unitary intertwiners $V_T,W_T$ for 
these representations $\pi_T(\A_S)$ and $\pi'_T(\A_{\bar S})$ on $\H_T$
. That is $V_T, W_T:\H_0\rightarrow \H_T$ satisfying
\begin{equation}
    \pi_T(a)V_T=V_T a\;\;\;\;\pi'_T(a')W_T=W_T a'\;\;\;\;\forall a\in \A_S,a'\in\A_{\bar S}
\end{equation}
and they are generally defined
up to unitaries on $\mathcal{H}_0$
that fix these intertwiner relations. Physically this non-uniqueness is a result of our freedom to choose the gravitational dressing scheme.  
The following unitary plays an important role
$K_T = V_T^\dagger W_T : \mathcal{H}_0 \rightarrow \mathcal{H}_0$, and can be shown to generate
automorphisms of $\A_S, \A_{\bar S}$
for each $T$.

Now the area operator of the extremal surface generates a flow which shifts the kink angle but otherwise keeps the field configuration unchanged at the classical level. However, in quantum theory the action which creates a kink while leaving the matter field unchanged is singular and so does not act on the Hilbert space $\mathcal{H}$. This singularity is usually understood to be canceled by a compensating flow of one-sided density matrix $-\ln \bm\rho$ which has the same singular structure in the vicinity of the extremal surface. 
Therefore, operators with formal decomposition \eqref{formal decomposition} and the crossed product structure naturally emerge as we try to incorporate the area operator into the gravitational algebra.

Using the aforementioned intertwiners we can more rigorously characterize the the action of the area operator. The requirement that
$\mathcal{H}_T, \pi_T,\pi_T'$ arise
from the kinked $\Sigma_T$ can be understood as the statement
that the automorphisms generated by $K_T$ are inner equivalent to modular flow (for any choice of cyclic and separating $\psi$) by boost angle $T$. That is:
\begin{equation}
\label{inneru}
    K_T \Delta^{i T}_\psi
    = u'_T u_T
\end{equation}
for some one parameter family
of unitaries in $u_T \in\A_S, u'_T \in \A_{\bar S}$. The challenge, that we do not solve here, is to quantize QFT on $M_T$, show that $\mathcal{H}_T, \pi_T,\pi_T'$ exist
and satisfies the above properties. Presumably this is the very least achievable for free quantum fields. The long wormhole results in Section~\ref{sec:4.2} can be understood as solving this problem using the split property and the special case where $M_T = M_0$.

Given \eqref{inneru} we can use the freedom in $V_T,W_T$ to remove the unitaries:
$V_T \rightarrow V_T (u_T')^\dagger$
and $W_T \rightarrow W_T \sigma_\psi^T(u_T) $. Then define:
\begin{equation}
V \Psi(T) = \int d T \;V_T^\dagger \Psi(T) \otimes \left| T \right>,
\end{equation}
which is a unitary that sends $V: \mathcal{H}
\rightarrow \mathcal{H}_0 \otimes L_2(\mathbb{R})$ and satisfies:
\begin{equation}
(a \otimes 1) V \Psi(T)
= V \pi_T(a) \Psi(T)
\qquad 
\pi(a') V \Psi(T)
= V \pi'_T(a') \Psi(T)
\end{equation}
where $\pi(a')
= \Delta_\psi^{i \hat{T}}(a'\otimes 1)\Delta_\psi^{-i \hat{T}}$
showing that the basic ingredients that define the crossed product algebra are naturally present on $\mathcal{H}$. The usual physical
arguments then assign the crossed product algebra to $S$ and its commutant to $\bar{S}$.

\section{Split property and the operator-valued weight}
\label{sec:ovw}
Before we dive into the discussion of crossed product algebras in gravity, we will first review the split property in quantum field theory \cite{DLsplit84}, and the notion of operator-valued weights (OVWs) \cite{HAAGERUP1979a, HAAGERUP1979b} for von Neumann algebra inclusions. Then we will show that from the split property, OVWs can be proven to exist quite generally for algebra inclusions corresponding to nested spacetime regions in QFT. We discuss this in two scenarios: 1) there is a finite gap between the boundaries of the two nested regions, which we call split inclusions; 2) the two nested regions share some common boundary, which we call contact inclusions. While for the first case, we prove that OVWs exist, for the second case, we show the existence of OVWs in special cases, but we expect that OVWs exist more generally. 

\subsection{Split Property}
We review the split property in quantum field theory, following \cite{DLsplit84}. For a nice review on split property, see Section 7.1 of \cite{Dutta:2019gen}. 

Assume that we have a QFT with a Hilbert space $\mathcal H$ and a cyclic and separating state $\ket{\psi}\in \H$. There are two regions $A$ and $B$ separated by a finite gap, with $\mathcal A_A$ and $\mathcal A_B$ the type III$_1$ von Neumann algebras for each region respectively. Split property says there exists a type I factor $\mathcal{R}$ between the two type III algebras:
\begin{equation}
    \A_A \subset \mathcal{R} \subset \A_B'.
\end{equation}
QFTs have the split property when the so-called nuclearity condition holds \cite{Buchholz:1986dy,GenNuclearity}. In this paper we will always assume the split property to hold for QFTs on the curved spacetimes of interest. However note that when the boundaries of regions $A$ and $B$ extend to infinity the split property can fail \cite{Witten:2018zxz}. We will not discuss this situation for most of this paper until Section \ref{sec:discussion}. When the standard split property holds, there is an isomorphism implemented by a unitary $U:\mathcal H\rightarrow \mathcal H\otimes\mathcal H$ which satisfies
\begin{equation}
    U(\mathcal A_A\vee \mathcal A_B) U^{-1}=\mathcal A_A\otimes \mathcal A_B.
\end{equation}
There is then a intermediate type I factor $\mathcal R$ which satisfies
\begin{equation}
    U\mathcal R U^{-1}= B(\mathcal H)\otimes I.
\end{equation}
There is, in fact, a canonical way to construct this algebra given some specific state on $\mathcal H$. Let $\ket{\psi}$ be a cyclic separating state we have chosen, there is a natural state $\ket{\psi}\otimes\ket{\psi}$ in the split space, and clearly this state is cyclic and separating for the algebra $\mathcal A_A\otimes\mathcal A_B$. But this is not the state we are going to use. Instead, we define a state $\ket{\xi}$ with the following property
\begin{equation}
    \mel{\xi}{a\otimes b}{\xi}=\mel{\psi}{ab}{\psi}
\end{equation}
where $a$ and $b$ are from $\mathcal A_A$ and $\mathcal A_B$ respectively. Note that this condition is non-trivial as in general $\mel{\psi}{ab}{\psi}$ does not factorize into $\mel{\psi}{a}{\psi}\mel{\psi}{b}{\psi}$. We can see $\ket{\xi}$ as a canonical way to purify the mixed state induced on $A\cup B$. Note that, since $\ket{\psi}\otimes\ket{\psi}$ is a cyclic separating state for $M=\mathcal A_A\otimes \mathcal A_B$, we have the following positive cone 
\begin{equation}
    \mathcal P=\overline{\Delta^{\frac{1}{4}}M_+\ket{\psi}\otimes\ket{\psi}}
\end{equation}
where $\Delta$ is the modular operator for the state $|\psi\rangle \otimes |\psi\rangle$. On the other hand, we also have a modular conjugation $J$ for this state. So now the set $\{M,\mathcal H\otimes\mathcal H,J,\mathcal P\}$ furnishes a standard form of the von Neumann algebra $M$, and from the theory of standard representations we know that $\ket{\xi}$ must be in the cone $\mathcal P$. So this will be the state we work with from now on. It can be shown that the modular conjugation operator $J_\xi=J_{\psi\otimes \psi}$ where $J_{\psi\otimes \psi}$ is the modular conjugation constructed with the state $\ket{\psi}\otimes\ket{\psi}$ (But there is no guarantee that the modular operator is also the same). 

In fact, the split discussed above is just a canonical way to define the split Hilbert space and algebra which is far from unique. It will be useful for our purpose to consider other split Hilbert spaces, and we will use the following theorem, which will be useful in our following discussion of algebras in python's lunch.
\begin{theorem}
    Let $\theta$ be a *-isomorphism between $\mathcal M\subset B(\mathcal H_1)$ and $\mathcal N\subset  B(\mathcal H_2)$, if there are cyclic separating vectors $\xi_1\in\mathcal H_1$ and $\xi_2\in\mathcal H_2$, then there is a unitary $U$ from $\mathcal H_1$ to $\mathcal H_2$ such that $ Ux U^{-1}=\theta(x)$ for any $x\in\mathcal M$.   
\end{theorem}

\subsection{Operator-valued weights}
\label{sec:3.2}
Next we introduce the notion of operator-valued weight and its properties, following \cite{HAAGERUP1979a,HAAGERUP1979b}. When we have von Neumann algebras $\N \subset \M$, an operator-valued weight $T$ is a map from the positive operators $\mathcal{M}_+$ of $\mathcal{M}$ to the extended positive part of $\hat{\mathcal{N}}_+$\footnote{The extended positive part of an algebra $\M$ is the set of all (possibly unbounded) lower semi-continuous, positive, densely defined operators on a Hilbert space $\H$ that are affiliated with $\M$.} of $\mathcal{N}$ that satisfies 
\begin{enumerate}
    \item $T(\lambda x) = \lambda T(x),\quad x\in \mathcal{M}_+,\, \lambda \geq 0$;
    \item $T(x+y)=T(x)+T(y),\quad x,y\in \mathcal{M}_+$;
    \item $T(a^* x a)=a^* T(x) a,\quad x\in \mathcal{M}_+, a\in \mathcal{N}$.
\end{enumerate}
Operator-valued weight is a generalization of the concept of conditional expectation, which is a map $E: \mathcal{M} \to \mathcal{N}$ with the properties of OVW and also satisfies $E(1)=1$. Normality (or lower semi-continuity), semi-finiteness and faithfulness are defined in a similar way as for usual weights. In the following when we say operator-valued weights we always mean n.s.f. operator-valued weights unless stated otherwise.

For finite dimensional quantum systems, given an operator $O\in \M$, the action of an OVW can be thought of as $\tr_{\N^c}(\rho_{\N^c} O)$, where $\N^c \equiv \N' \wedge \M$, the complement of $\N$ in $\M$, and $\rho_{\N^c}$ is an operator supported only on $\N^c$. If $\rho_{\N^c}$ is normalized, we will get a conditional expectation.

For infinite dimensional systems, conditional expectation and OVW are the appropriate notions of the ``tracing out'' procedure, although it is often the case that $\mathcal{N}^c$ is trivial and so the degrees of freedom we trace out involve the
operators that extend $\mathcal{N}$ to $\M$. Conditional expectations have been very useful in studying subjects like bulk reconstruction \cite{Faulkner:2020hzi}, global symmetries in quantum systems \cite{Casini:2019kex}, etc. However, they do not generally exist for inclusions of type III algebras, which is the case for nested regions in quantum field theory. In such a scenario, as we will see, OVW is the relevant notion. In this paper, we will use $P(\M,\N)$ to denote the set of OVWs from $\M$ to $\N$. We will also use $P(\M)$ to denote the set of weights on $\M$. Here we review some important results in the theory of OVW \cite{HAAGERUP1979a,HAAGERUP1979b}, which we will refer to in the remaining of the paper. 

First of all, for $\N\subset \M$, OVWs do not change the modular flow generated by weights on $\N$:
\begin{theorem}\label{thm:ovwflow}
    Let $T\in P(\mathcal{M},\mathcal{N})$, and for a given weight $\omega$ on $\N$,
\begin{equation}
    \sigma_t^{\omega\circ T}(x)=\sigma_t^{\omega}(x),\quad x\in \mathcal{N}.
\end{equation}
\end{theorem}
\noindent Furthermore, such a property on modular flow also guarantees the existence of an OVW:
\begin{theorem}\label{thm2}
    Let $\M,\N$ be von Neumann algebras, and $\N \subseteq \M$. Let $\omega_1,\omega_2$ be normal, faithful, semifinite (n.f.s.) weights on $\N$ and $\M$ respectively. If $\sigma^{\omega_2}_t(x)=\sigma^{\omega_1}_t(x)$ for any $x\in \N$, then there exists a unique n.f.s. OVW such that $\omega_2 = \omega_1 \circ T$.
\end{theorem}

As we see above, for finite-dimensional quantum systems, OVWs correspond to (possibly unnormalized) density matrices in the relative commutant $\N^c \equiv \N' \wedge \M$. Analagously, for general von Neumann algebra inclusions, OVWs give rise to well-defined modular flows on the relative commutant:
\begin{theorem}\label{thm3}
    Let $T\in P(\M,\N)$. For any $\omega\in P(\N)$, $\sigma^{\omega\circ T}_t(\N^c)=\N^c$, and the restriction of $\sigma^{\omega\circ T}$ to $\N^c$ is independent of the choice of $\omega$. 
\end{theorem}
\noindent For later convenience, we just use $\sigma^T_t$ to denote the restriction of $\sigma^{\omega\circ T}_t$ to $\N^c$.

From the above theorems, we see that, when OVWs exist, they give rise to weights like $\omega \circ T$ that seem like a ``tensor product'' between $\N^c$ and $\N$, since its modular flow keeps $\N^c$ and $\N$ fixed simultaneously.

When there is an OVW $T: \M \to \N$, there exists a dual OVW $T^{-1}:\N' \to M'$. This is given by the following lemma
\begin{lemma}\label{dual weight}
    Let $\mathcal N\subseteq\mathcal M$ be two von Neumann algebras, and $P(\mathcal M,\mathcal N)$ be the set of operator-valued weights from $\mathcal M$ to $\mathcal N$, then we have
    \begin{equation}
        P(\mathcal M,\mathcal N)\neq\varnothing\Leftrightarrow P(\mathcal N',\mathcal M')\neq\varnothing. 
    \end{equation}
\end{lemma}
\noindent The modular flows are related by the following theorem
\begin{theorem}
    Let $\M,\N$ be von Neumann algebras on a Hilbert space $\H$. There exists a bijection $\alpha$ of $P(\M,\N)$ onto $P(\N',\M')$ such that $\sigma^{\alpha(T)}_t=\sigma^T_{-t}$.
\end{theorem}
\noindent This theorem gives the intuition that OVW corresponds to an invertible density matrix in the relative commutant. However, this might be a trivial statement when the relative commutant is trivial. In the special case where the OVW is a conditional expectation, the dual is still generally an OVW. The dual is a conditional expectation only when the index for the inclusion is finite. 

The existence of OVW implies a subalgebra structure for the crossed product algebra. For an algebraic inclusion $\N \subset \M$, when there is a weight $\omega$ on $\N$, one can obtain a weight on $\M$ using the OVW $T$: $\omega_T=\omega \circ T$. From theorem \ref{thm:ovwflow}, we know that the modular flow of $\omega_T$ and $\omega$ agree on $\N$. Therefore, the crossed products have a inclusion structure, and also an OVW $\widetilde T$ between them. We can draw a commutative diagram
\begin{equation}
    \begin{tikzcd}
        \M \arrow[r, "T"] \arrow[d, "P",leftarrow]
        & \N \arrow[d, "P",leftarrow]\\
       \M \rtimes_\sigma \mathbb{R} \arrow[r, "\widetilde T",rightarrow]& \N \rtimes_\sigma \mathbb{R}
    \end{tikzcd}
\end{equation}
where $P$ is the operator-valued weight from the type II$_\infty$ algebra to the type III$_1$ algebra. 

Lastly, we discuss how the existence of OVW depends on the types of von Neumann algebras $\M$ and $\N$. 
\begin{lemma}
\label{thm:types}
    Let $\mathcal N\subseteq\mathcal M$ be an inclusion of von Neumann algebras, then there exist an operator-valued weight from $\mathcal M$ to $\mathcal N$ in any one of the following cases: 
    \begin{enumerate}
    \setlength{\itemindent}{1em}
        \item $\mathcal N$ and $\mathcal M$ are semifinite;
        \item $\mathcal N$ is a sum of type I factors;
        \item $\mathcal M$ is a sum of type I factors.
    \end{enumerate}
\end{lemma}
For a von Neumann algebra inclusion $\mathcal{N}\subseteq \mathcal{M}$, there is no normal conditional expectations from $\mathcal{M}$ to $\mathcal{N}$ when $\mathrm{type}(\mathcal{N})>\mathrm{type}(\mathcal{M})$. However, operator-valued weights exist more generally.

\subsection{Operator-valued weights for split inclusions}
\begin{figure}
    \centering
    \begin{subfigure}[b]{0.45\textwidth}
        \includegraphics[width=\textwidth]{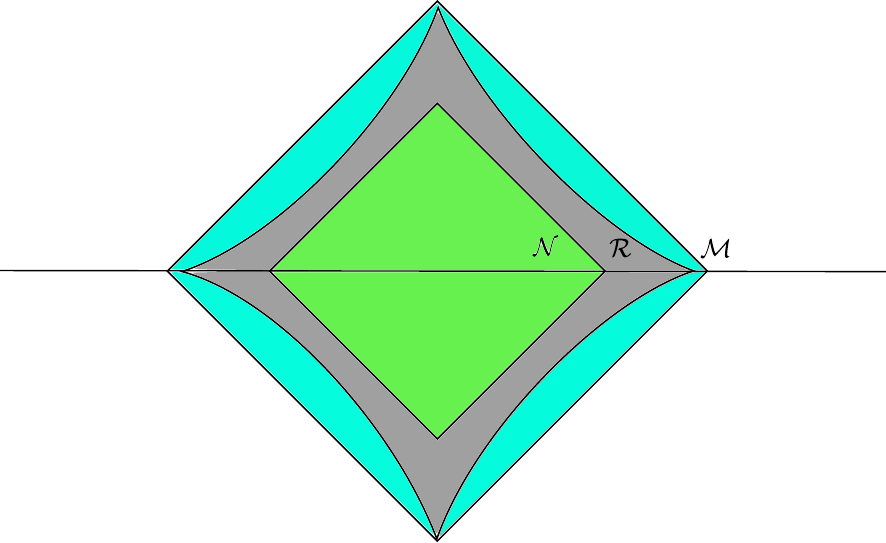}
        \caption{}
        \label{split inclusion}
    \end{subfigure}
    \hfill
    \begin{subfigure}[b]{0.45\textwidth}
        \includegraphics[width=\textwidth]{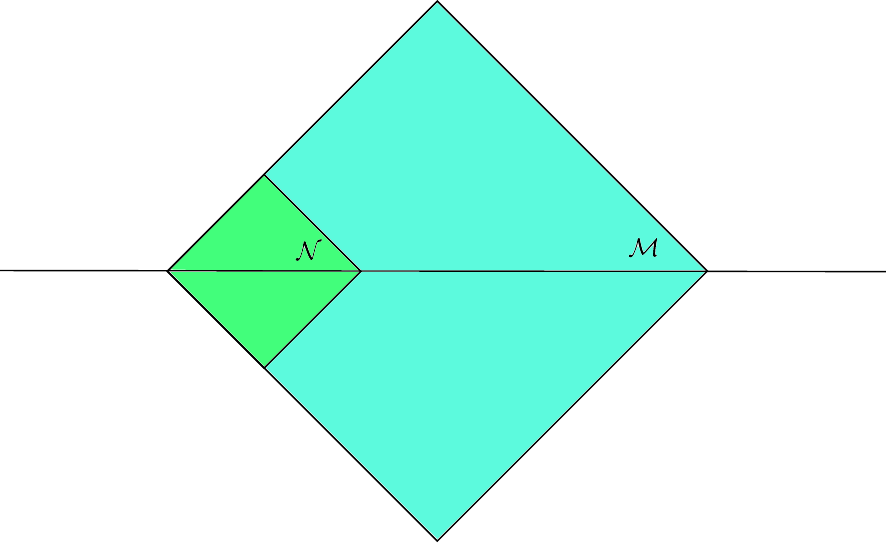}
        \caption{}
        \label{contact inclusion}
    \end{subfigure}
    \caption{An example of split inclusion (left) and contact inclusion (right). For the split inclusion, there is a type I factor $\mathcal{R}$ such that $\N\subset\mathcal{R}\subset \M$. Note that $\mathcal{R}$ is not associated to any geometric region despite how it is shown.}
\end{figure}

First we discuss an inclusion for nested regions without any shared boundaries, which we call split inclusion, for example the one shown in Figure~\ref{split inclusion}. In this case, we have split property for $\N$ and $\M'$, which we can use to construct an operator-valued weight $T: \M \to \N $. This construction
is well known, see for example \cite{Faulkner:2024gst}. First note that after the split mapping, there exists a type I factor $\mathcal{R}$ such that
\begin{equation}
    U \M U^{-1}=\mathcal{B}(\mathcal{H}) \otimes \M, \quad U \mathcal{R} U^{-1}=\mathcal{B}(\mathcal{H}) \otimes 1.
\end{equation}
So we can construct a conditional expectation $E_1: \M \to \mathcal{R}$ by simply applying a state $\omega_1$ on the second factor:
\begin{equation}
\label{condexp}
    E_1 = \mathrm{Ad}_{U^{-1}} \circ ( \mathbb{1}\otimes \omega_1) \circ \mathrm{Ad}_U.
\end{equation}

Next, according to Lemma \ref{thm:types}, there is always an operator-valued weight from a type I algebra to a type III algebra. So we can find
\begin{equation}
    T_2: \mathcal{R} \to \N.
\end{equation}
Given a faithful normal state $\omega_2$ on $\N$, then $\omega_2 \circ T_2$
is a weight on $\mathcal{R}$.
Since $\mathcal{R}$ is type-I such a weight always has the form:
\begin{equation}
\omega_2 \circ T_2(\cdot) = {\rm Tr}_{\mathcal{R}} D(\cdot)
\end{equation}
for some possibly unbounded $D$
affiliated to $\mathcal{R}$.
In fact this defines the so-called
spatial derivative $D = \Delta(\omega_2 / T_2^{-1})$ where $T_2^{-1}$ is a weight
on $\mathcal{N}'$ \cite{ConnesSpatial}. This is a cousin of the relative modular operator for weights. 
Now we reveal the exceptional natural of the modular flow for $\omega_2 \circ T_2$:
it is determined by a density matrix $D^{is}(\cdot) D^{-is}$ where $D$
itself is a relative modular operator. 
Hence modular flow fixes, as a set, the relative
commutant of this inclusion $\mathcal{R} \wedge \mathcal{N}'$,
simply because that is what relative modular operators do. At the same time the flow on this set is ``opposite'' to the usual modular flow.

Finally the desired operator-valued weight is constructed from composing the two: $T=T_2 \circ E_1$.
The construction above is helpful to understand the modular flow of a weight of the form $\omega_2 \circ T$. After the split, this weight just becomes $\omega_2 \circ T_2\tp \omega_1$. Therefore the modular operator also takes a factorized form
\begin{equation}
    \D_{\eta_{\omega_2 \circ T}}=D \tp \D_{\eta_{\omega_1}}
\end{equation}
where $\eta_{\omega_2 \circ T}$ and $\eta_{\omega_1}$ are the vector representations of the weights $\omega_2 \circ T$ and $\omega_1$ respectively. The modular flow on $\M$ is then given by
\begin{equation}
    \s^{\omega_2\circ T}_t=\mathrm{Ad}_{U^{-1}} \circ \s^{\omega_2\circ T_2\tp \omega_1}_t\circ \mathrm{Ad}_U.
\end{equation}
This form of the the modular flow suggests that the flow induced by the operator-valued weight $T$ on the relative commutant $\M\wedge\N'$ can be thought as having a factorized structure, which is the tensor product of a modular flow on $\M \wedge \mathcal R$ and another one on $\mathcal{R}\wedge \N'$.

\subsection{Operator-valued weights for contact inclusions}
\label{sec:3.4}
Next we discuss the existence of operator-valued weights between two nested regions with shared boundaries, which we call contact inclusions. See for example Figure~\ref{contact inclusion}.
In this case, the above argument does not work since there is no split between $\N$ and $\M'$. In this case there is not a gap between the subalgebra $\mathcal N$ and the large algebra $\mathcal M$, so it is not clear whether there is a operator-valued weight. 

\begin{figure}[h!]
  \centering
  \includegraphics[width=0.6\textwidth]{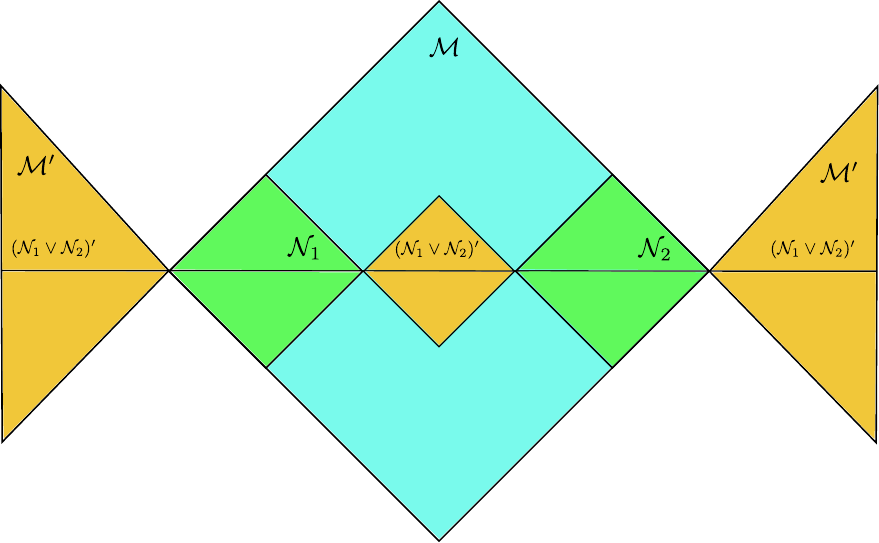} 
  \caption{Inclusion of algebras $\N_1 \subset \N_1\vee \N_2 \subset \M$ in (1+1)-dimensions. We use the split property to show that there exists an operator-valued weight from $\M$ to $\N_1\vee \N_2$ and therefore from $\M$ to $\N_1$. }
  \label{two subregions contact inclusion}
\end{figure}

However, as we will show, there are special cases where the OVW still exists. First we look at (1+1)-dimensions, there exists a operator-valued weight $T$ from $\mathcal M$ to $\mathcal N$. To do this, we first consider the configuration in Figure~\ref{two subregions contact inclusion}, in this case we have two connected subregions with subalgebras $\mathcal N_1$ and $\mathcal N_2$ respectively.   
We will denote $\mathcal N=\mathcal N_1\vee\mathcal N_2$, which is the smallest von Neumann algebra generated by $\mathcal N_1$ and $\mathcal N_2$. We start by looking at the commutants of these algebras, the orange regions in the figure denotes the commutant $\mathcal N'$, which is a disconnected region, while the two wedges on the left and right denotes the commutant $\mathcal M'$, and we have $\mathcal M'\subseteq\mathcal N'$. However, notice that, although $\mathcal N$ and $\mathcal M$ are in contact, the inclusion $\M'\subseteq\N'$ can be handled by using the fact that $\N' \wedge \M \subset \M$ is split.
Under the split isomorphism for this inclusion we have $\mathcal{M}' \vee (\N' \wedge \M) \cong \mathcal{M}' \otimes (\N' \wedge \M)$ and the inclusion of interest $\mathcal{M}' \subset \mathcal{N}'$
is isomorphic to $\mathcal{M}'\otimes 1 \subset \mathcal{M}' \otimes (\N' \wedge \M)$. Then we can construct a faithful normal conditional expectation for this inclusion in the same way as in \eqref{condexp}.
 Since there is a conditional expectation from $\mathcal N'$ to $\mathcal M'$
\noindent we can immediately conclude that there is a operator-valued weight $T$ from $\mathcal M$ to $\mathcal N=\mathcal N_1\vee\mathcal N_2$. Now note that $\mathcal N$ contains two disconnected subregions, and we can use the split inclusion again to see that there is a operator-valued weight, in fact a conditional expectation $E$, from $\mathcal N$ to $\mathcal N_1$. Again using the composition of operator-valued weights, we get the conclusion that there is a operator-valued weight $T_1=E\circ T$ from $\mathcal M$ to $\mathcal N_1$. Thus we have proved the existence of an operator-valued weight from the larger subregion to the smaller one in the situation of a contact inclusion.

\begin{figure}[h!]
  \centering
  \includegraphics[width=0.5\textwidth]{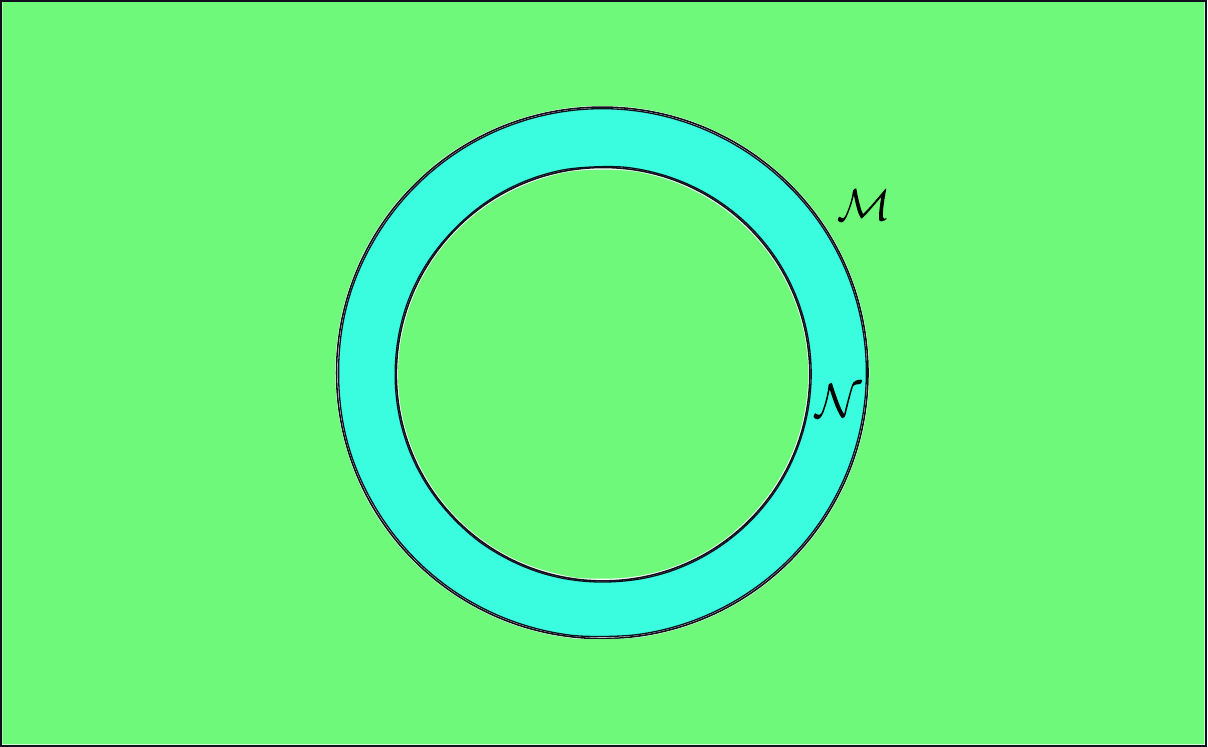} 
  \caption{An algebraic inclusion for QFT in at least (2+1) dimensions. $\M$ is the algebra for the region inside the outer circle, while $\N$ is the algebra for the annulus region. For simplicity, we have only shown a time slice.}
  \label{2d contact inclusion}
\end{figure}

The argument above can be generalized to higher dimensions for certain contact inclusions. For example we can consider the (2+1)-dimensional inclusion as shown in Figure~\ref{2d contact inclusion}, where the algebra $\mathcal M$ is the region within the outer circle, while the subalgebra $\mathcal N$ is for the annulus region between the two circles. 
Using the same commutant argument as above, notice that the commutant of $\mathcal N$ again contains two disconnected pieces, we can similarly show that there is an operator-valued weight from $\mathcal M$ to $\mathcal N$. This argument also applies to higher dimensions. In this higher dimensional setting one question that we do not know the answer to: does there exist operator-valued weight when $\mathcal N$ and $\mathcal M$ only touch on a segment of the outer boundary but not the whole.

\section{Gravitational algebras with two areas}
\label{sec:pythonslunch}
In this section, we first review some basic facts about long wormholes, after which we construct gravitational algebras in different regions of such a spacetime when both extremal-surface areas are included.

\subsection{Long wormholes and the python's lunch}
As mentioned in the introduction, this kind of geometry can be formed with a heavy thin shell behind the horizons \cite{Goel:2018ubv,Balasubramanian:2022gmo}, and this will be the spacetime we study\footnote{However, see \cite{Bak:2021qbo} for an example where the long wormhole can be formed with a matter field in JT gravity.}; see Figure \ref{fig:PL-Penrose} again. The boundary dual of this is known as the partially entangled thermal state (PETS) \cite{Goel:2018ubv}, which can be formed from insertion of a heavy operator $\cal O$ followed by Euclidean time evolution on both sides:
\begin{equation}
\label{eq:PETS}
    |\Psi\rangle = \frac{1}{\sqrt{Z}}\sum_{ij} e^{-\beta_L E_i/2-\beta_R E_j/2}\mathcal{O}_{ij}|E_i\rangle_L |E_j\rangle_R.
\end{equation}
where $\mathcal{O}_{ij}$ are the matrix components of $\mathcal{O}$ in the energy eigenbasis. The bulk geometry is then prepared by the corresponding Euclidean gravitational path integral. The Lorentzian geometry has two bifurcation horizons that are locally minimal extremal surfaces, and the left and right regions are diffeomorphic to the exterior regions of AdS Schwarzschild black holes with inverse temperatures $\beta_1$ and $\beta_2$ respectively. Note that $\beta_1$, $\beta_2$ are in general different from $\beta_L$, $\beta_R$ in equation \eqref{eq:PETS}, due to the heavy operator insertion.

As a side note, this bulk saddle only dominates the gravitational path integral when $\beta_L,\beta_R $ are small enough. When they are larger than a critical value, another saddle with trivial extremal surfaces will dominate \cite{Antonini:2023hdh}, whose Lorentzian section contains two thermal AdS spaces and a closed universe. This is the analog of Hawking-Page transition with a thin shell present.

\begin{figure}[h!]
  \centering
\includegraphics[width=0.35\textwidth]{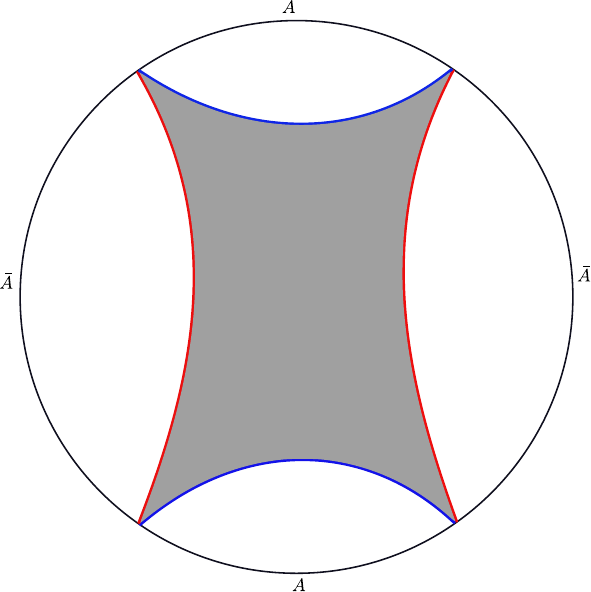} 
  \caption{The two candidate extremal surfaces (red and blue) for boundary subregions $A$, which is a union of two connected subregions. The python's lunch is the region shaded in gray.}
  \label{fig:PL-Vacuum AdS}
\end{figure}

The long wormhole geometry is an example of spacetimes that exhibit a python's lunch, since on any Cauchy slice that passes through the two bifurcate horizons, the area of the transverse directions reaches local minima at the bifurcate horizons, and reaches local maximum at the so-called bulge surface located between them. There are also examples of python's lunch when the extremal surfaces are non-compact. Figure~\ref{fig:PL-Vacuum AdS} shows the two candidate extremal surfaces for the union of two boundary subregions in vacuum AdS. In this case, there is a non-trivial bulge surface which crosses on itself \cite{Arora:2024edk}. With quantum matter present, the python's lunch can appear when we consider generalized entropies and quantum extremal surfaces. One important example of this is an evaporating black hole \cite{Penington:2019npb,Almheiri:2019psf}. We now return to the discussion of the long wormhole, but we will comment more on general geometries with python's lunch in Section \ref{sec:discussion}.

For a long wormhole, we will specifically focus on a subspace of the classical phase space. This subspace is spanned by $A_1,A_2,T_1,T_2$, the areas of the two extremal surfaces and the kink angles (related to time shifts) that are conjugates to the areas. Pick any Cauchy slice $\Sigma$ that passes through the two extremal surfaces, after the action of the area operators, we have a Cauchy slice $\Sigma_{T_1,T_2}$ where the extremal surfaces have kink angles $T_1,T_2$. We can then evolve $\Sigma_{T_1,T_2}$ to obtain the full solution $M_{T_1,T_2}$. As discussed in \ref{subsec:generalextremal}, when there exists a Cauchy slice $\widetilde \Sigma_{T_1,T_2}$ in the original spacetime $M$ that is diffeomorphic to $\Sigma_{T_1,T_2}$, $M_{T_1,T_2}$ will also be diffeomorphic to $M$. We can see this is exactly the case for a long wormhole, due to the isometries in the exterior regions. For any Cauchy slice $\Sigma$, one can simply do one-sided boosts (Schwarzschild time translations) in the exterior regions to obtain the slice $\widetilde \Sigma_{T_1,T_2}$ that has the same Cauchy data as $\Sigma$ except for the kinks at the extremal surfaces. Therefore, when we put quantum fields on the long wormhole, the Hilbert spaces are naturally identified, and the gravitational Hilbert space, when taking into account the two extremal surface areas, is
\begin{equation}
    \H_{QFT}\otimes L^2(\mathbb{R}_{X_1})\otimes L^2(\mathbb{R}_{X_2}).
\end{equation}

In the long wormhole spacetime, we label the left bifurcate horizon $\gamma_1$, and the right bifurcate horizon $\gamma_2$. For any Cauchy slice that passes through the two bifurcate horizons, it is naturally divided into three parts: the region to the left of $\gamma_1$, the region between $\gamma_1$ and $\gamma_2$, and the region to the right of $\gamma_2$. We denote them as $L$, $M$, and $R$, respectively. We are also interested in the union of the above three regions, and we will use shorthands for them: $MR=M\cup R$, $LM=L\cup M$, and $LR=L\cup R$. These region labels will be used to denote the algebras that are associated with their causal development. For a region $Y$, we use $\A_Y$ for its QFT algebra, and $\widetilde \A_Y$ for its gravitational algebra.

\subsection{Algebras with two areas: split property at work}\label{sec:4.2}
We now study the assignment of gravitational algebras to different regions in a long wormhole. We first consider the algebras when areas of both extremal surfaces are included. This is the case when we take the microcanonical ensemble in both ADM Hamiltonians $H_L$, $H_R$, where they both have $O(1)$ fluctuations. In the $G_N \to 0$ limit, the relevant degrees of freedoms are the fluctuations $X_1=\beta_1(H_L-\la H_L\ra)$, $X_2=\beta_2(H_R-\la H_R\ra)$. 

Physically, observers in each region should have access to the areas of the extremal surfaces which bound the subregion. Therefore, regulated versions of these area operators should be added into the gravitational algebras. Moreover, time shifts, or equivalently kink angles at extremal surfaces, are accessible to observers in a subregion which contain an extremal surface in its interior instead of on its boundary. One example is the region $L\cup M$ in Figure~\ref{fig:PL-Penrose}. Finally, we also expect these gravitational algebras to be complete, in the sense that algebras of complementary regions should be commutants of each other, which is also referred to as the Haag duality \cite{haag2012local}. 

We now start from the gravitational algebra of the python's lunch in the middle. In Section~\ref{sec:review} we emphasized the fact that the area operator, strictly speaking, is not an operator but only a bilinear form as it has diverging fluctuation. The resolution to this problem is to add another `operator' with the correct divergence structure which cancels the divergence in the area operator, namely one which locally approximates a one-sided boost. It has been argued that under reasonable assumptions the modular flow of an arbitrary state restricted to a subregion locally approaches a boost near its boundary (see e.g. \cite{Jensen:2023yxy,Sorce:2024zme}). However, we cannot simply combine the area operators with the one-sided modular Hamiltonian induced on the middle region by some state. The reason is that these one-sided modular Hamiltonians contain singularities in the vicinity of both extremal surfaces, but our goal is to regularize the area operator of \emph{a single} surface. Another guess would be to add, for example, the one-sided modular Hamiltonian induced by a state on the union $M\cup R$, which has the correct singularity structure in the neighborhood of $\g_1$. But the problem is that this one-sided modular flow is not localized within the middle region and does not preserve the algebra $\mathcal \A_M$, which cannot be used to construct an algebra localized within $M$. 

Now recall that the split property discussed above allows for a factorization of the QFT Hilbert space into left and right factors. Intuitively, this factorization provides a way to define an `operator' which is localized in the middle while being singular at only one of the extremal surfaces. This is done by choosing an operator which acts as an one-side modular flow on one of the factors while trivially on the other.

\begin{figure}[t]
  \centering
  \includegraphics[width=0.8\textwidth]{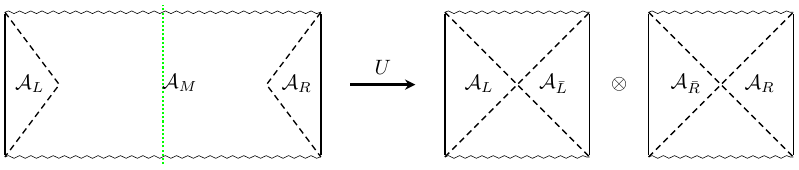} 
  \caption{The split map $U$ factorizes the quantum field theory on a long wormhole into the tensor product of two quantum field theories defined on two two-sided black holes.}
  \label{split python}
\end{figure}

Using the split property introduced in Section~\ref{sec:ovw}, we can factorize the quantum field theory Hilbert space $\H_{QFT}$ into $\H_1\otimes\H_2$ using a unitary transformation $U$. We choose $\H_1$ and $\H_2$ to be the QFT Hilbert spaces on two-sided Schwarzschild black holes with inverse temperatures $\beta_1$ and $\beta_2$, respectively.\footnote{In fact, it is possible to choose other split Hilbert spaces, which is just saying there are different ways to choose the purification of algebras $\A_L$ and $\A_R$. As we will see the choice does not affect the following discussion. Furthermore, it can be proved that in fact all such `canonical purifications' are unitarily equivalent. This is a result of the theory of standard forms of von Neumann algebras, which states that the standard form of a von Neumann algebra is unique up to unitary transformations \cite{takesaki2003theory}.} See Figure~\ref{split python}.

After the split map, we are able to pick a reference state with the form of a tensor product $|\eta\ra \equiv |\eta_1\ra \otimes |\eta_2\ra$ on $\H_1\otimes \H_2$, where $|\eta_1\ra$ and  $|\eta_2\ra$ are Hartle-Hawking states with inverse temperatures $\beta_1$ and $\beta_2$ respectively.

We now construct the gravitational algebra of the middle region with two areas included. On the split Hilbert space $\H=\H_1\otimes \H_2$, we can define the modular operator induced by $\ket{\eta_1}$ on $\A_{\bar L}$, which can be written as\footnote{For simplicity we will omit subscripts indicating algebras in modular and relative modular operators when the relevant algebra is clear from context.}
\begin{equation}
\Delta_{\eta_1}\tp \mathbb{1}_{\H_2}\equiv e^{-\til h_1}\tp\mathbb{1}_{\H_2}=e^{-(\til {\bm h}_{\bar L}-\til {\bm h}_L)}\tp\mathbb{1}_{\H_2}
\end{equation}
Note that this operator acts trivially on the tensor factor $\H_2$, as indicated by the factor $\mathbb{1}_{\H_2}$. In the following without causing confusion we will drop the trivial factor and write it as $\D_{\eta_1}=e^{-\til h_1}$. $\til h_1$ is a well-defined operator on the split Hilbert space, so it can be transformed back to $\H_{QFT}$ as $h_1=U^\dagger \til h_1 U$. Since the split map $U$ preserves the locality of operators in the exterior regions, $h_1$ becomes a (inverse) boost on $\A_L$, and therefore the following operator 
\begin{equation}\label{reg area 1}
    X_1+h_1
\end{equation}
is the regularized area operator for $\g_1$, which is the sum of two well-defined operators. For the factor $\H_2$ we repeat the procedure above to get the operator
\begin{equation}\label{reg area 2}
    X_2+h_2
\end{equation}
where $h_2$ is the modular Hamiltonian induced by the state $\ket{\eta_2}$ on $\A_{\bar R}$. Note that $X_{1,2}$ does not act on $\H_{QFT}$ and are therefore unaffected by the split map $U$. So we can map~\eqref{reg area 1}, \eqref{reg area 2} to the split Hilbert space to get $X_1+\til h_1$ and $X_2+\til h_2$. In the following we will mainly work with the split Hilbert space as expressions are simpler to write down here.

Now we construct the crossed product algebra for $\A_{M}$, or equivalently $\A_{\bar L}\tp\A_{\bar R}$ after the split map. In the split Hilbert space we can construct the crossed products for each tensor factor independently. For $\H_1$ we take the crossed product to be
\begin{equation}
    \widetilde {\A}_{\bar L}=\{\A_{\bar{L}}, X_1+\til h_1\}''\equiv\A_{\barL}\rtimes_\sigma\mathbb{R}_{X_1}
\end{equation}
which is a type $\text{II}_\infty$ factor. $\widetilde \A_{\bar L}$ acts on $\H_1\tp L^2(\mathbb R)$, where the $L^2(\mathbb R)$ factor is acted by multiplicatively by $X_1$. A tracial weight can be readily defined for $\til \A_{\bar L}$ as
\begin{equation}
    \ket{\t_1}=\int dX_1\;e^{\frac{X_1}{2}}\ket{\eta_1}\tp\ket{X_1}
\end{equation}
Similarly we can obtain a crossed product $\widetilde \A_{\barR}=\A_{\barR}\rtimes_\sigma\mathbb{R}_{X_2}$ for the tensor factor $\H_2$. 

Now we take their tensor product, which is again a type $\text{II}_\infty$ von Neumann factor, with a trace taking a tensor product form
\begin{equation}
    |\t_1\rangle \otimes |\t_2\rangle=\int dX_1dX_2\;e^{\frac{X_1+X_2}{2}}\ket{\eta_1}\tp\ket{\eta_2}\tp\ket{X_1}\tp\ket{X_2}.
\end{equation}
The type II$_\infty$ gravitational algebra in the middle region in the original Hilbert space is obtained by implementing the unitary $U$
\begin{equation}
    \widetilde \A_M=U^\dagger\big(\widetilde \A_{\barL}\tp\widetilde \A_{\barR}\big)U
    =\{\A_{M},X_1+{h}_1,X_2+{h}_2\}'',
\end{equation}
and the trace is given by
\begin{equation}
    \ket{\t}=U^{-1} (|\t_1\ra \otimes |\t_2\ra)=\int dX_1dX_2\;e^{\frac{X_1+X_2}{2}}U^\dagger\big(\ket{\eta_1}\tp\ket{\eta_2}\big)\tp\ket{X_1}\tp\ket{X_2}.
\end{equation}

It should be pointed out that despite $\widetilde \A_M$ is a well-defined gravitational algebra in the middle region where observables are properly dressed to the extremal surfaces, it does not have a clear microscopic origin on the boundary, as it is not the entanglement wedge of any boundary subregions. The boundary interpretation of gravitational algebras in the python's lunch will be left to future works. 

Next we discuss the gravitational algebras assigned to other regions. First, the gravitational algebra in the union $L\cup R$ is given by
\begin{equation}
    \widetilde \A_{L\cup R}=\{e^{i\Pi_1 h_1}\A_{L}e^{-i\Pi_1 h_1},e^{i\Pi_2 h_2}\A_{R}e^{-i\Pi_2 h_2},X_1,X_2\}''
\end{equation}
which is the commutant of $\widetilde \A_{M}$.

The gravitational algebras for the left and right exteriors are respectively
\begin{equation}
    \til \A_L=\{e^{i\Pi_1 h_1}\A_{L}e^{-i\Pi_1 h_1},X_1\}'',\quad \A_R=\{e^{i\Pi_2h_2}\A_{R}e^{-i\Pi_2 h_2},X_2\}''.
\end{equation}
Obviously $\til \A_L$, $\til\A_R$ and $\til \A_{L\cup R}$ are all type II$_\infty$ factors. Now we discuss the gravitational algebras for regions $M\cup R$ and $L\cup M$. We have
\begin{equation}
\label{eq:AMR}
    \til\A_{MR}=\til \A_{L}'=\{\A_{MR}, X_1+ h_1,X_2,\Pi_2\}'' = \{\A_{MR}, X_1+ h_1\}''\otimes B(L^2(\mathbb{R}_{X_2})),
\end{equation}
\begin{equation}
\label{eq:ALM}
    \til\A_{LM}=\til \A_{R}'=\{\A_{LM}, X_2+ h_2,X_1,\Pi_1\}'' =\{\A_{LM}, X_2+ h_2\}''\otimes B(L^2(\mathbb{R}_{X_1})).
\end{equation}
Notice that these algebras not only contain $X$'s but also $\Pi$'s, so they contain a subalgebra $B(L^2(\mathbb R))$. This in fact makes sense: we discussed in Section~\ref{sec:review} that $\Pi_{1,2}$ should be interpreted as the kink angle of the Cauchy slice at surfaces $\g_{1,2}$. So if we are considering the subregion $M\cup R$ for example, we should have access to the kink angle at $\g_2$. Thus $\Pi_2$ is in the gravitational algebra for $M\cup R$. Clearly all these gravitational algebras are also of type II$_\infty$ as von Neumann types are preserved under taking commutants.

Another question is that the form of the algebras \eqref{eq:AMR} and \eqref{eq:ALM} seems to be in contradiction to the expectation that one should get, for example
\begin{equation}
    \til\A_{MR}=\til \A_M\vee\til\A_R=\{\A_M,e^{i\Pi_2 h_2}\A_{R}e^{-i\Pi_2 h_2},X_1+ h_1,h_2,X_2, \Pi_2\}''.
\end{equation}
However, this contradiction is only prima facie. In fact, $\Pi_2$ can be generated from subalgebras $U^\dagger \A_{\bar R}U\subset \A_{M}$, $X_2$, $h_2$ and $e^{i\Pi_2 h_2}\A_{R}e^{-i\Pi_2 h_2}$ in the same way as in the case of a two-sided black hole. It then turns out that the two expressions for $\til \A_{MR}$ are equivalent. From the discussion above we conclude that gravitational algebras we defined satisfy completeness and additivity, as algebras of unions are unions of algebras, and algebras of complements are commutants of algebras. As we will see in Section \ref{sec:singlearea}, these properties could break down if we only take area sum or difference into consideration.

Before we move on, we would like to comment on another way to define the gravitational algebra in a finite spacetime region, that is to introduce an observer as discussed in~\cite{Chandrasekaran:2022cip,Jensen:2023yxy}, where they also used a crossed product construction. There due to a projection to positive energy states of the observer we end up with a type II$_1$ von Neumann algebra. In our case, in contrast, we interpret our construction as gravitational dressing to extremal surfaces instead of an observer, which leads to a type II$_\infty$ algebra.

\subsection*{Generalized entropy in python's lunch}
Having defined the trace on the algebra, we can now calculate the type $\text{II}_\infty$ entropy defined on $\widetilde \A_M$. We will see that this entropy is just the generalized entropy for the middle region as it can be written as a formal sum of the quantum field theory entropy and the total area which bounds in the middle region. As before, we consider the classical-quantum state given by
\begin{equation}
\ket*{\Ps}=\int_{-\infty}^{\infty} d X_1d X_2\, g_1(X_1)g_2(X_2)\ket{\ps}\tp\ket{X_1}\tp\ket{X_2},
\end{equation}
where $g_1(X_1)g_2(X_2)=\epsilon_1^{1 / 2}\epsilon_2^{1 / 2} \mathcal{G}_1(\epsilon_1 X_1)\mathcal{G}_2(\epsilon_2 X_2)$ with $\e_1,\e_2\ll 1$. We have also assumed that there is no entanglement between the two areas. Since $\widetilde \A_M$ is isomorphic to $\widetilde \A_{\bar{L}} \otimes \widetilde \A_{\bar R}$, computing the entropy of $|\Psi\ra$ for the algebra $\widetilde \A_M$ is equivalent to computing entropy of $U|\Psi\ra$ for the algebra $\widetilde \A_{\bar{L}} \otimes \widetilde \A_{\bar R}$.

The approximate density matrix of $U|\Psi\ra$ for the algebra $\widetilde \A_{\bar{L}} \otimes \widetilde \A_{\bar R}$ is given by
\begin{equation}
    \rho_{U\Ps}= |g_{1}(X_1)g_2(X_2)|^2e^{-X_1-X_2}\Delta_{U\ps,\eta}
\end{equation}
where the relative modular operator $\Delta_{U\ps,\eta}$ refers to the algebra $\A_{\bar{L}} \otimes \A_{\bar R}$.

From this form of the density matrix we can derive the type $\text{II}_\infty$ entropy
\begin{equation}
\begin{aligned}
    S_M&=-\expval{\ln\rho_{\Ps}}{\Ps}=-\expval{\ln\rho_{U\Ps}}{U\Ps}\\
    &=\la X_1 +X_2 \ra -\la U\Psi|\ln \Delta_{U\psi,\eta} |U\Psi\ra -\int dX_1|g_1(X_1)|^2\ln|g_1(X_1)|^2-\int dX_2|g_2(X_2)|^2\ln|g_2(X_2)|^2,
\end{aligned}
\end{equation}
where $\la X_i \ra = \int dX_i|g_i(X_i)|^2 X_i$. 

Similar to the calculation for two-sided black holes, we use some formal decomposition to show that the above entropy agrees with the generalized entropy in the middle region.

First of all, we can formally write the QFT relative modular operator $\Delta_{U\psi,\eta}$ as tensor products of ``density matrices'' 
\begin{equation}
\Delta_{U\ps,\eta}=\bm{\rho}^{\barL\barR}_{U\ps}\tp(\bm{\rho}_{\eta_1}^{L})^{-1}\otimes (\bm{\rho}_{\eta_2}^{R})^{-1},
\end{equation}
whose logarithm gives
\begin{equation}
    \ln \Delta_{U\ps,\eta}=\ln \bm\rho^{\barL\barR}_{U\ps}+(\bm h_L+\bm h_R)
\end{equation}
where $\bm h_L$ and $\bm h_R$ are the one-sided boost operators in left and right exteriors for the corresponding Hartle-Hawking state. We will also assume that $U$ factorizes into the formal decomposition $U=\bm U_L\otimes \bm U_{M}\otimes \bm U_R$, so $\expval{\ln \bm \rho_{ U \ps}^{\barL\barR}}{ U \ps} = \la \psi|\ln \bm\rho_{ \ps}^{M}|\psi\ra$. Then combining the above and using formal decomposition $X_{1,2}=\frac{\bm {\delta A}_{1,2}}{4G}+\bm h_{L,R}$, we find the entropy is formally
\begin{align}
    S_M\approx \expval{\frac{\bm{\delta A}_1+\bm{\delta A}_2}{4G}}_\Ps-\expval{\ln \bm \rho_{ \ps}^{M}}{ \ps}-\int dX_1|g_1(X_1)|^2\ln|g_1(X_1)|^2-\int dX_2|g_2(X_2)|^2\ln|g_2(X_2)|^2
\end{align}
which agrees with the generalized entropy in the middle region, together with the entropy from the two area fluctuations.

Again, although the algebraic entropy for the python's lunch region is a well-defined quantity in the bulk theory, it is not manifestly a microscopic entropy, since the python's lunch is not an entanglement wedge for any boundary region. 

For gravitational algebras in regions $L$ and $R$, it is not hard to compute the entropies of the classical-quantum state $|\Psi\ra$, and find that they agree with the corresponding generalized entropies upon formal decomposition. In fact, up to the split map, the calculation will be identical to that of \cite{Chandrasekaran:2022eqq}. We will refrain ourselves from repeating and only list the results
\begin{equation}
\begin{aligned}
    S_{L,R} &=\la X_{1,2} \ra -\la U\Psi|\ln \Delta_{U\psi,\eta} |U\Psi\ra -\int dX_{1,2}|g_{1,2}(X_{1,2})|^2\ln|g_{1,2}(X_{1,2})|^2\\
    &\approx \expval{\frac{\bm{\delta A}_{1,2}}{4G}}_\Ps-\expval{\ln \bm \rho_{ \ps}^{L,R}}{ \ps}-\int dX_{1,2}|g_{1,2}(X_{1,2})|^2\ln|g_{1,2}(X_{1,2})|^2.
\end{aligned}
\end{equation}

\section{Gravitational algebras with only the sum or difference of areas}
\label{sec:singlearea}
In the above discussion of gravitational algebras and Hilbert spaces, $X_1$ and $X_2$ are two independent modes with $O(1)$ fluctuations. Each of them contains an area operators of an extremal surface that is either $\bm{\d A_1}$ or $\bm{\d A_2}$. Alternatively, one can pick the two modes to be two independent linear combinations of $X_1$ and $X_2$, for example, $X_+\equiv X_1+X_2$ and $X_-\equiv X_1-X_2$, each of which has $O(1)$ fluctuations. 

\begin{figure}[t]
    \centering
    \begin{subfigure}[b]{0.3\textwidth}
        \centering
        \includegraphics[width=\textwidth]{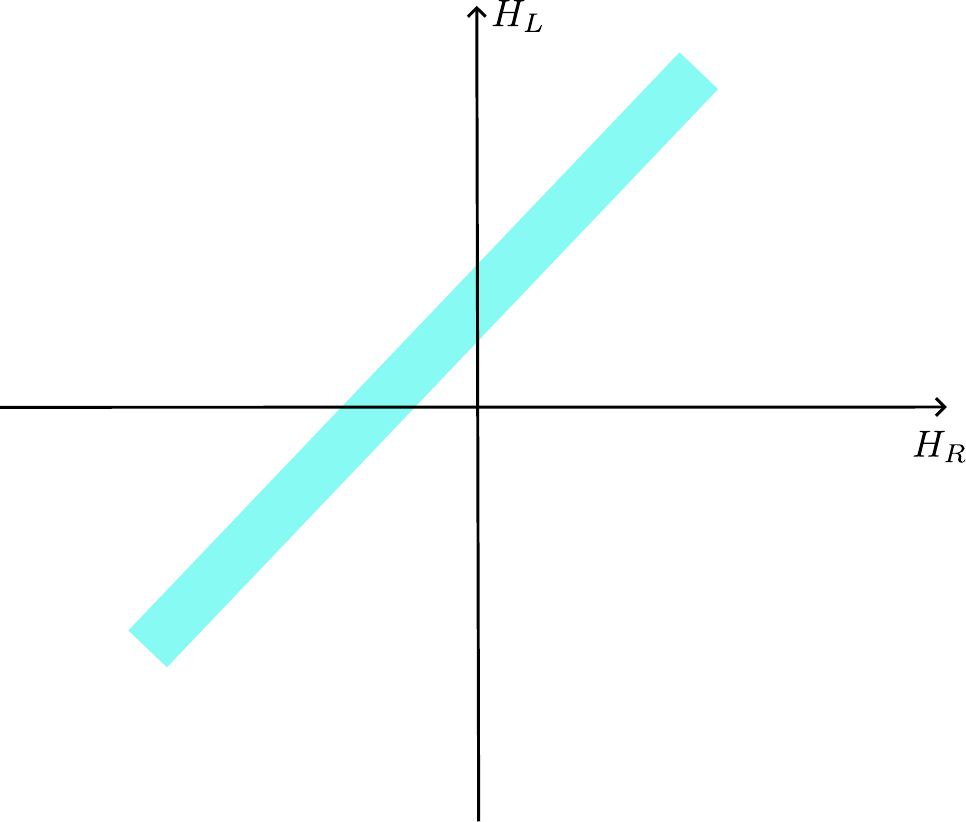}
        \label{fig:sub2}
    \end{subfigure}
    \hfill
    \begin{subfigure}[b]{0.3\textwidth}
        \centering
        \includegraphics[width=\textwidth]{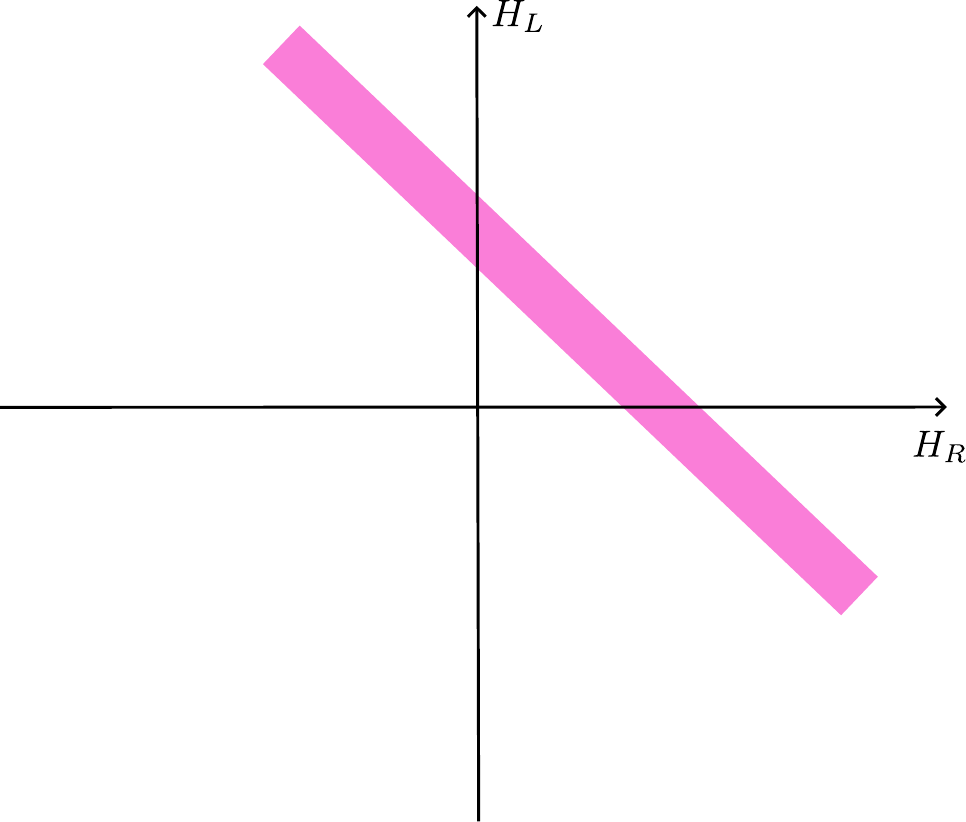}
        \label{fig:sub3}
    \end{subfigure}
    \hfill
    \begin{subfigure}[b]{0.3\textwidth}
        \centering
        \includegraphics[width=\textwidth]{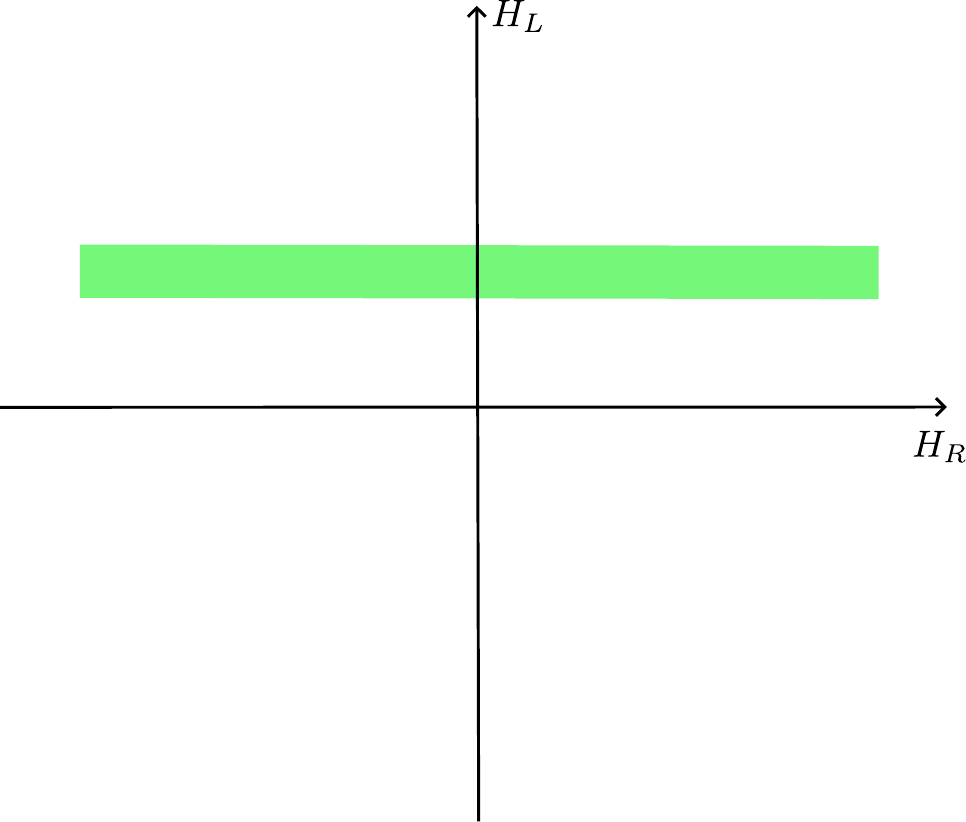}
        \label{fig:sub1}
    \end{subfigure}

    \caption{
Microcanonical windows in the $(H_L,H_R)$ plane. From left to right,
$H_L-H_R$, $H_L+H_R$, and $H_L$ are restricted to $O(1)$ windows.
The conjugate variable to the restricted combination consequently has
$O(1)$ fluctuations. Along the unconstrained direction, the corresponding
combination of Hamiltonians fluctuates by $O(N)$, so its conjugate has
$O(1/N)$ fluctuations and becomes fixed in the strict large-$N$ limit.
}
    \label{fig:spec}
\end{figure}

However, in certain situations, it would be interesting to consider the microcanonical ensemble in only $X_+$ or $X_-$. The microcanonical ensemble for $X_1+X_2$ is relevant when the total energy of the two CFTs is constrained. In this ensemble, the mode $X_+$ has $O(1)$ fluctuations, while $X_-$ is left with $O(N)$ fluctuations. In terms of the conjugate variables, the fluctuation of $\Pi_+=(\Pi_1+\Pi_2)/2$ is $O(1)$, and the fluctuation of $\Pi_-=(\Pi_1-\Pi_2)/2$ is $O(1/N)$. Therefore, in the strict large $N$ limit, $X_-/N$ becomes a center for all gravitational algebras. While $X_+$ and $\Pi_+$ can be well-defined operators in gravitational algebras, $\Pi_-$ will be a fixed number. See Figure \ref{fig:spec}. 

For the microcanonical ensemble of the difference $X_1-X_2$, the analysis for fluctuations of operators is similar to the above, with $+$ and $-$ exchanged. This ensemble is useful for studying the phase transition in entanglement wedges.  Furthermore, a single area operator may not be well-defined in general geometries. For example, in the geometry of Figure \ref{fig:PL-Vacuum AdS}, due to the IR divergence at the asymptotic boundary, the areas of the extremal surfaces are divergent, but the area difference is still well-defined. 

In this section we will consider both cases. We will see that the area sum is straightforward while the area difference operator needs more elaboration. We should keep in mind that these algebras always contain a center which factors out, and we can put them back in the end. For certain entropies the center might contribute the dominant term (scaling as $1/G_N^{1/2}$) to the entropy, in which case the contribution we compute below should be understood as the subleading terms. However for the main entropies of interest (such as $S_M$ in the area sum case, or the difference of entropies in the area difference case) this dominant term cancels and so our answers are in fact the leading contribution.

\subsection{Gravitational algebras with the area sum}

Now we consider the case where only the area sum mode is available in the gravitational algebras. There are several reasons to start from this case. First it is the simplest example and by studying it we can get some intuition which can be generalized to more complicated cases. Secondly, we expect that the generalized entropy in the python's lunch region will be encoded in the gravitational algebra with only the area sum mode, as the generalized entropy only knows the \emph{total} area of the extremal surfaces bounding the subregion.  

We start with the algebra in the middle region. Formally, we again need to combine the area sum operator $\frac{\bm{\delta A_1+\delta A_2}}{4G}$ with some one-sided modular Hamiltonian to make it well-defined. From the previous discussion we know that the following combination is a well defined operator in the middle region
\begin{equation}
    \frac{\bm{\delta A_1}+\bm{\delta A_2}}{4G}+U^\dagger (\bm h_{\barL}+\bm h_{\bar R}) U=X_1+X_2+h_1+h_2\equiv X_++h_+.
\end{equation}
Therefore for the middle region we assign the following crossed product algebra
\begin{equation}
    \widetilde \A_{M}^+=\{\A_M,X_++h_+\}''
\end{equation}
which is clearly a type II$_\infty$ factor. By taking the commutant of $\til \A_M^+$ we can get the algebra for the union $L\cup R$, which is also type II$_\infty$
\begin{equation}
    \widetilde \A_{LR}^+ = \{e^{i\Pi_+ h_+} (\A_L\vee \A_R) e^{-i\Pi_+ h_+}, X_+ \}''.
\end{equation}
However, if we have access to only the left or right exterior, we are not able to recover $X_+$, since only $X_1$ or $X_2$ is accessible. Therefore, their gravitational algebras are given by 
\begin{equation}
    \widetilde \A_L^+ = e^{i\Pi_+ h_+} \A_L e^{-i\Pi_+ h_+}
\end{equation}
\begin{equation}
    \widetilde \A_R^+ = e^{i\Pi_+ h_+} \A_R e^{-i\Pi_+ h_+}.
\end{equation}
It can be easily seen that the additivity of gravitational algebras breaks down in this case as we have
\begin{equation}
   \widetilde \A_{LR}^+ \neq  \widetilde \A_L^+ \vee \widetilde \A_R^+.
\end{equation}
This is a not a surprising result. In fact, by projection to fixed $t_L-t_R$ we are introducing some non-local constraints on the geometry which correlates the kink angles at $\gamma_1$ and $\gamma_2$, resulting in non-local observables in $L\cup R$ which cannot be generated by considering $L$ or $R$ individually. Violations of algebraic additivity can be found in~\cite{Casini:2019kex,Casini:2020rgj,Casini:2021zgr,Shao:2025mfj} in the context of quantum field theory with global symmetries. See also \cite{Leutheusser:2024yvf} for the discussion of the breakdown of additivity for generalized free fields.

The gravitational algebra for $M\cup R$ can be found to be
\begin{equation}
    \begin{aligned}
    \label{eq:AMR+}
        \widetilde \A_{MR}^+&=(\widetilde \A_{L}^+)'= \{e^{i\Pi_+ h_+} \A_{MR} e^{-i\Pi_+ h_+} , X_+ + h_+, \Pi_+ \}''\\
&=e^{i\Pi_+ h_+}\{ \A_{MR}  , X_+, \Pi_+ \}''e^{-i\Pi_+ h_+}\\
&=e^{i\Pi_+ h_+}\left( \A_{MR} \otimes  B(L^2 ({\mathbb{R}_{X_+}})) \right) e^{-i\Pi_+ h_+}.
    \end{aligned}
\end{equation}
Careful readers may find that the commutant of $\widetilde \A_L^+$ can also be written as $\{ \A_{MR}  , X_+ + h_+, \Pi_+ \}$. However, this algebra can be shown to be equivalent to the first line of \ref{eq:AMR+}, whose proof can be found in the proof of the Takesaki duality \cite{takesaki1973duality}. Similarly, the gravitational algebra in $L\cup M$ region can be found to be
\begin{equation}
       \widetilde \A_{LM}^+=(\widetilde \A_{R}^+)'=e^{i\Pi_+ h_+}\left( \A_{LM}  \otimes B(L^2(\mathbb{R}_{X_+})) \right) e^{-i\Pi_+ h_+}.
\end{equation}

In contrast to the previous crossed-product type II$_\infty$ algebras, it is easy to see gravitational algebras $\til \A_{L}^+$, $\til \A_{R}^+$, $\til \A_{LM}^+$, $\til \A_{MR}^+$ are all type III$_1$ factors. This is a result of the projection we have done. Another interesting feature is that $\til \A_{LM,MR}^+$ contain all the bounded operators $B(L^2(\mathbb R))$ acting on the $L^2(\mathbb R)$ factor of the Hilbert space. This is consistent with our expectation from a physical perspective. As we are fixing the kink angle difference $t_L-t_R$, accessing one of them amounts to accessing both. Therefore $\Pi_+\equiv t_L+t_R$ can be obtained on $M\cup R$ or $L\cup M$. The algebras defined above clearly satisfy the following inclusions 
\begin{equation}\label{chain of inclusion}
    \til \A_L^+,\til \A_R^+\subset \til \A_{LR}^+\;\;\;\;\;\;\til \A_L^+,\til\A_M^+\subset \til \A_{LM}^+\;\;\;\;\;\;\til \A_R^+,\til\A_M^+\subset \til \A_{MR}^+.
\end{equation}
Due to the type III nature of the above algebras, it is not possible to define trace and entropy. However, we can compute the algebraic entropy on the only type II$_\infty$ algebra $\widetilde A_M^+$. This algebra is a crossed product and it is again easier to do the calculation in the split Hilbert space, where the trace is 
\begin{equation}
    \ket{ \t_+}=\int dX_+ \;e^{\frac{X_+}{2}}\ket{\eta}\tp\ket{X_+}
\end{equation}
with $\ket\eta=\ket {\eta_1}\tp\ket{\eta_2}$ as before. And a semiclassical state in this case can be written as 
\begin{equation}
    \ket{\Ps}=\int dX_+ \e^{\frac{1}{2}}g(\epsilon X_+)\ket{\ps}\tp\ket{X_+}.
\end{equation}
Repeating the entropy calculation discussed previously, we find that the entropy of this state on $\til\A_M^+$ is given by\footnote{In principle there should be an extra term accounting for the fluctuation of the center variable, which in this case is $X_-$. We will drop this term as it only contributes a constant in all entropy calculations. }
\begin{equation}
    S_M=\expval{\frac{\bm{\delta A_1+\delta A_2}}{4G}}_\Ps-\expval{\ln \bm\rho_\ps^{M}}{\ps}-\int dX_+|g(X_+)|^2\ln|g(X_+)|^2.
\end{equation}
That is, except for a different contribution from area fluctuations, the algebra $\widetilde \A_{M}^+$ in fact encode the generalized entropy defined in the middle region. This agrees with our intuition that the generalized entropy is the sum of the quantum field theory entropy and the \emph{total} area of the extremal surface which bound the subregion. It should therefore be enough to include only the area sum mode to get the generalized entropy for the subregion. 

\subsection{Gravitational algebras with the area difference: Use of operator-valued weights}
\label{sec:5.2}

In order to add the area difference mode, we need to find some one-sided modular flow in the middle region that matches the divergence in $\frac{\bm {\d{A_1}}-\bm {\d A_2}}{4G}$. Applied to the algebra in the middle region, the generator we need should approach a boost that goes up in the vicinity of $\g_1$, while it should look like an inverse boost near $\g_2$. It seems hard, or maybe impossible, to find such a modular flow using a state on $\A_M$, since generally the modular flow of such a state should be future-directed near both extremal surfaces \cite{Sorce:2024zme}. However, as we will see, we can find such an operator which generates the correct flow structure with the help of the operator-valued weights discussed in Section~\ref{sec:ovw}.

\begin{figure}
    \centering
    \includegraphics[width=0.5\linewidth]{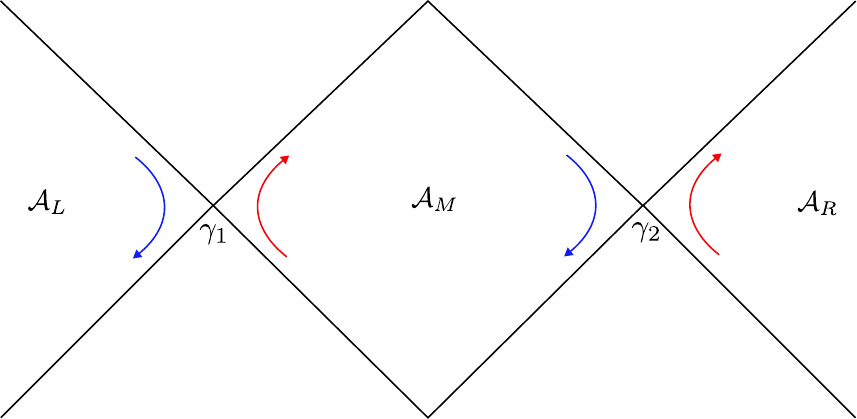}
    \caption{The flow directions near the extremal surfaces from the relative modular operator $\D_{\omega_2\circ T,\w_1}$. While the flow is purely geometric in regions $L$ and $R$, it is only geometric near the edges in the region $M$. These directions of the flow makes it a well-defined operator when combined with $A_1-A_2$.}
    \label{fig:ovw_flow}
\end{figure}
From discussion in Section~\ref{sec:ovw}, there exists an operator-valued weight $T$ from $\A_M\vee \A_R= \A_{MR}$ to $\A_R$. Given any state $\omega$ on $\A_R$, we can consider a weight $\omega\circ T$ defined on $\A_{MR}$. From Theorem~\ref{thm:ovwflow}, we know that  $\sigma_t^{\omega\circ T}(x)=\sigma_t^{\omega}(x)$ for $x\in \A_R$. In other words, the modular flow $\sigma_t^{\omega\circ T}(\cdot)$ preserves $\A_R$ and thus the relative commutant $\A_M$. For simplicity, we choose $\omega$ to be a Hartle-Hawking state $\omega_2$ (as a linear functional) on $\A_R$.

Now we consider the relative modular operator $\Delta_{\omega_2\circ T,\w_1}$\footnote{
Relative modular operators
can be defined with respect to a weight $\omega_2 \circ T$ as in \cite{ArakiMasuda}. Consider the purification of the state $\w_1$ for $\A_L$, on $\H_{QFT}$, denoted as $\ket{\eta_1}$. This is always possible as $\H_{QFT}$ furnishes a standard form of $\A_L$. 
Using the semi-cyclic representation of $\omega_2 \circ T$, denoted $\eta_{\omega_2 \circ T}(\cdot)$, one can then construct the relative modular data in the usual way, where the Tomita operator $S$ acts between the two Hilbert spaces, $S n \eta_1 = \eta_{\omega_2 \circ T}(n^\dagger)$
for $n$ in an appropriate subset of $\mathcal{A}_{MR}$. Then $\Delta_{\omega_2\circ T,\w_1} = S^\dagger S$.
}, where $\omega_1$ is the Hartle-Hawking state on $\A_L$. We define $h_- \equiv -\ln \Delta_{\omega_2\circ T,\w_1}$ for later convenience. The flow generated by $\Delta_{\omega_2\circ T,\w_1}$ agrees with the modular flow induced by $\omega_2\circ T$ on $\A_{MR}$ while it becomes the reversed modular flow induced by $\w_1$ on $\A_L$. By construction, when restricted to $\A_R$, it also agrees with the modular flow induced by $\w_2$. Hence the flow it generates should approach local two-sided boosts in the vicinity of the two extremal surfaces, as depicted in Figure~\ref{fig:ovw_flow}.  
Note that this flow, when restricted to $\A_M$, has the desired structure to regularize the area difference operator $\frac{\bm {\d A_1}-\bm{\d A_2}}{4G}$ as it flows in opposite directions in the vicinity of $\g_1$ and $\g_2$. In fact, the restriction of the flow on $\A_M$ is determined purely by the operator-valued weight as shown in Theorem~\ref{thm3}. 

It is worth mentioning that there is a dual of the above construction. For a given operator-value weight $T:\A_{MR}\rightarrow \A_R$, there is a dual weight $T^{-1}:\A_{LM}\rightarrow \A_L$ guaranteed by Lemma~\ref{dual weight}. We can then define the relative modular operator $\D_{\w_1\circ T^{-1},\w_2}=\D^{-1}_{\w_2\circ T,\w_1}$ with respect to $\A_{LM}$, which generates the inverse flow. In what follows we will stick to $\Delta_{\omega_2\circ T,\w_1}$ but readers should keep in mind that all constructions can be done in a dual way.  

An intuitive, but mathematically non-rigorous, way to understand the flow generated by $\Delta_{\omega_2\circ T,\w_1}$ is to assume a tensor factorization between regions $M$ and $R$, so that the operator-valued weight can then be formally written as
\begin{equation}\label{eq:5.6}
    T(x)=\tr_M(\bm\sigma x)\otimes \mathbb{1}_M
\end{equation}
which is a partial trace of the operator on $M$ with a non-singular density matrix $\bm\sigma$ inserted on $M$. The flow restricted to $\A_M$ can be then viewed as being generated by a modular Hamiltonian $-\ln\bm\sigma$, which is an intrinsic feature of the operator-valued weight and therefore independent of the choice of $\w_2$. Similar to one-sided boosts, $-\ln\bm\sigma$ is not a well defined operator, as it is singular in the vicinity of $\g_1$ and $\g_2$. 

One is then tempted to construct a crossed product algebra in the middle region by adding $X_-+h_-$, since formally we have
\begin{equation}\label{reg diff area}
    X_-+h_- = \frac{\bm{\delta A_1-\delta A_2}}{4G}-\ln(\bm \sigma)
\end{equation}
for some ``density matrix'' associated to $T$ in the middle region. Such a crossed product algebra can be constructed as
\begin{equation}
    \widetilde \A_M^- = \{ \A_M, X_- + h_- \}''.
\end{equation}
Since $h_-$ generates a modular flow restricted to $\A_M$, this crossed algebra is unitarily equivalent to the usual modular crossed product and hence is of type II$_{\infty}$.

Once we have the gravitational algebra in the middle region, gravitational algebras assigned to the other regions can be found in a similar way as for the area sum case discussed above:
\begin{equation}\label{gravity algebra 1}
    \widetilde \A_{LR}^- = (\widetilde \A_M^-)'= \{ e^{i\Pi_- h_-}(\A_L \vee \A_R)e^{-i\Pi_- h_-}, X_- \}'',
\end{equation}
\begin{equation}\label{gravity algebra 2}
     \widetilde \A_{LM}^-=e^{i\Pi_- h_-}\left( \A_{LM} \otimes B(L^2 (\mathbb{R}_{X_-})) \right)e^{-i\Pi_- h_-},\quad \widetilde \A_{MR}^-=e^{i\Pi_- h_-}\left( \A_{MR} \otimes B(L^2 (\mathbb{R}_{X_-})) \right)e^{-i\Pi_- h_-},
\end{equation}
\begin{equation}\label{gravity algebra 3}
    \widetilde \A_L^- = e^{i\Pi_- h_-} \A_L e^{-i\Pi_- h_-},\quad \widetilde \A_R^- = e^{i\Pi_- h_-} \A_R e^{-i\Pi_- h_-}.
\end{equation}
The types of these von Neumann algebra are the same as in the area sum case. And clearly they satisfy the same inclusions~\eqref{chain of inclusion}.

Our next goal is to give an expression for the generalized entropy difference using these algebras. In particular if we consider our previous results in Section~\ref{sec:4.2} for the generalized entropies calculated using the algebras with both $\bm{\d A_1}$ and $\bm{\d A_2}$ and consider a limiting state where the area sum fluctuations become large, we find a finite answer for the entropy difference. We expect this new algebra to correctly account for this limit and indeed we will see this is the case by explicit calculation. 

Naively we should relate this generalized entropy difference to a difference type-II entropies, as in the usual crossed product, however physically the relevant algebras are no longer type II. Instead we will use a difference between relative entropies on gravitational algebras $\til \A_{MR}^-$ and $\til \A_R^-$. It is well known that conditional entropies often have well defined expressions even for type III von Neumann algebras, at least in the presence of a conditional expectation \cite{gao2020relative}. As usual one proceeds by using relative entropies. Here we won't have a conditional expectation but in its place we will use a special operator-valued weight that we construct below.

For simplicity we will do the calculation in the twirled representation, which is obtained by conjugating the gravitational algebras in~\eqref{gravity algebra 1}~\eqref{gravity algebra 2}~\eqref{gravity algebra 3} by a unitary transformation $e^{-i\Pi_- h_-}$. The advantage of this representation is that the relevant algebras $\til\A_{MR}^-$ and $\til \A_R^-$ take a tensor factorized form
\begin{equation}
    \widetilde \A_{MR}^- = \A_{MR} \otimes B(L^2 (\mathbb{R})),
\end{equation}
\begin{equation}
    \widetilde \A_{R}^- = \A_{R} \otimes \mathbb{1}\,, \quad\widetilde \A_{L}^- = \A_{L} \otimes \mathbb{1}.
\end{equation}

Note that there is a chain of algebra inclusion 
\begin{equation}
    \widetilde \A_{M}^-\subset (\widetilde \A_{M}^- \vee \widetilde \A_{R}^-) \subset \widetilde \A_{MR}^-
\end{equation}
where for the first inclusion, there is a non-trivial relative commutant $\widetilde \A_R^-$, and for the second inclusion, the relative commutant is trivial.

The classical-quantum state for which we are doing the entropy calculation takes the following form in the twirled representation
\begin{equation}
\label{twirling}
    | \Xi\rangle = e^{-i\Pi h_-}|\hat \Xi\rangle= e^{-i\Pi h_-}\left(|\psi\rangle \otimes \int dX_- g(X_-)|X_-\ra \right)
\end{equation}
where we use $\ket*{\hat\Xi}$ to denote the classical-quantum state in the untwirled representation, where it takes a factorized form. 
Restricting $\Xi$ to $\widetilde \A_{MR}^-$ defines the linear functional state
$\widetilde{\omega}_{\Xi}(\cdot) = \omega_\Xi|_{\til\A_{MR}^-}(\cdot)$.
Moreover, we introduce the following weight defined on $\widetilde \A_{MR}^-$:
\begin{equation}
\widetilde{\omega}_{T}(\cdot) = \w_2 \circ T(\cdot) \otimes \Tr(e^{X_-}(\cdot) ) 
\end{equation}
where we used the usual trace
on the $L^2(\mathbb{R})$ Hilbert space. 
Note that $\widetilde{\omega}_{T}$ can be written as the state $\omega_2$ composed with an operator-valued weight $\widetilde T$
\begin{equation}
   \widetilde{\omega}_{T} = \omega_2 \circ \widetilde T,
\end{equation}
where $\tilde T$ is given by:
\begin{equation}
\widetilde{T} = T \otimes \Tr(e^{X_-}\cdot )  :\widetilde \A_{MR}^- \rightarrow\widetilde \A_{R}^-.
\end{equation}

It is possible to show that $\widetilde{\omega}_{T}$ restricts to a semifinite faithful tracial weight, i.e. the trace, on the type-II$_\infty$ algebra of $\widetilde \A_M^-$, and we will comment on the significance of this weight below. We also define the Hartle-Hawking state $\widetilde{\omega}_{1}$ on $\widetilde \A_L^- = \mathcal{A}_L \otimes 1$ to be 
\begin{equation}
    \widetilde{\omega}_{1}(\cdot \otimes 1)=\w_{1}(\cdot)
\end{equation}
We start with the relative entropy between $\widetilde{\omega}_{\Xi}$ and $\widetilde{\omega}_{T}$ on $\widetilde \A_{MR}^-$,
\begin{align}
      S_{\rm rel}(\widetilde{\omega}_{\Xi}|\widetilde{\omega}_{T})_{MR} &= -\la \Xi| \ln \Delta_{\widetilde{\omega}_{T},\widetilde{\omega}_{\Xi}'} |\Xi \ra \nonumber \\
      &= -\la \Xi|\ln \Delta_{\widetilde{\omega}_{1},\widetilde{\omega}_{\Xi}}|\Xi\rangle - \la \Xi| \ln \Delta_{\widetilde{\omega}_{T},\widetilde{\omega}_{1}}|\Xi \ra \nonumber \\
      &= - \la \Xi|\ln \Delta_{\widetilde{\omega}_{1},\widetilde{\omega}_{\Xi}}|\Xi\ra - \la \Xi| \ln \Delta_{\w_2\circ T,\omega_1}+X_-|\Xi\ra
      \label{fiseterm}
\end{align}
where in the first line we use the Araki definition for the relative entropy, and we denoted $\tilde \A_{MR}^-$ and $\tilde \A_R^-$ as $MR$ and $R$ for simplicity. For this expression  we must use the state on the commutant
$\widetilde{\omega}_{\Xi}'= \omega_{\Xi}|_{\widetilde{\mathcal{A}}_L^-} $ in the definition of the relative modular operator.
To derive the second line, we first use the following cocycle relations\footnote{These cocycle relations sometimes require
extra support projections in the case that some of the weights/states involved are not faithful on their respective algebras, see for example \cite{Ceyhan:2018zfg}. We obviously have faithfulness for $\widetilde{\omega}_{1},\widetilde{\omega}_{T},\widetilde{\omega}_{\Xi}'$ if $\psi$ is sufficiently entangled, but this is subtle for $\widetilde{\omega}_{\Xi}$. Presumably sufficient entanglement generated by the twirling in \eqref{twirling} makes $\Xi$ cyclic and separating for $\til \A_{MR}^-$. In any case, including support projections is not usually an issue, and we don't bother tracking these here so that the calculation is easier to follow. }
\begin{equation}
    \Delta_{\widetilde{\omega}_{1},\widetilde{\omega}_{T}}^{is} \Delta_{\widetilde{\omega}_{1},\widetilde{\omega}_{\Xi}}^{-is} = \Delta_{\widetilde{\omega}_{\Xi}',\widetilde{\omega}_{T}}^{is}\Delta_{\widetilde{\omega}_{\Xi}',\widetilde{\omega}_{\Xi}}^{-is}\,\,,\quad \Delta_{\widetilde{\omega}_{\Xi}',\widetilde{\omega}_{T}} =\Delta_{\widetilde{\omega}_{T},\widetilde{\omega}_{\Xi}'}^{-1}\,\,,\quad  \Delta_{\widetilde{\omega}_{1},\widetilde{\omega}_{T}} =\Delta_{\widetilde{\omega}_{T},\widetilde{\omega}_{1}}^{-1}
\end{equation}
where recall that 
$\Delta_{\widetilde{\omega}_{\Xi}',\widetilde{\omega}_{\Xi}} =
\Delta_{\Xi;L}$ is the (non-relative) modular operator for $\Xi$ and $\widetilde{\A}^-_L$. 
This gives:
\begin{equation}
    \Delta_{{\widetilde{\omega}_{T}},{\widetilde{\omega}_{1}}}^{-is} \Delta_{{\widetilde{\omega}_{1}},\widetilde{\omega}_{\Xi}}^{-is} = \Delta_{{\widetilde{\omega}_{T}},\widetilde{\omega}_{\Xi}'}^{-is}\Delta_{\Xi;L}^{-is}
\end{equation}
then, assuming the $s$ derivative of the above equation exists\footnote{This is not always the case, but there should be a sufficiently general and nice class of states where the derivative exists. Such details will not be considered in this paper.}, we take the $s$ derivative of the above equation and take $s=0$ to get
\begin{equation}
    \ln \Delta_{{\widetilde{\omega}_{T}},\widetilde{\omega}_{\Xi}'} = -\ln \Delta_{\Xi;L} +\ln \Delta_{{\widetilde{\omega}_{T}},{\widetilde{\omega}_{1}}} +\ln \Delta_{{\widetilde{\omega}_{1}},\widetilde{\omega}_{\Xi}}.
\end{equation}
The first term on the right is killed inside the $\Xi$ expectations while the second term on the right is easy to compute as ${\widetilde{\omega}_{T}}$ takes a tensor product form.  This proves \eqref{fiseterm}.

Next we evaluate $- \la \Xi|\ln \Delta_{{\widetilde{\omega}_{1}},\widetilde{\omega}_{\Xi}}|\Xi\ra$. First note that the state $\Xi$ is a product state between the QFT Hilbert space and $L^2(\mathbb{R})$, entangled by the twirling factor $e^{-i\Pi h_-}$. It is not hard to find the restriction of this state to $\widetilde \A_L^-=\A_L\otimes \mathbb{1}$ by tracing out the $B(L^2(\mathbb{R}))$ factor:
\begin{equation}
   \rho_1(\cdot) \equiv \widetilde{\omega}_{\Xi}'(\cdot \otimes 1) = \int dp |\widetilde g(p)|^2 \omega_{\psi,L} \circ \Delta_{\w_1}^{ip} (\cdot) \Delta_{\w_1}^{-ip} 
\end{equation}
where we used the momentum basis for simplicity, and $\widetilde g(p)$ is the Fourier transform of $g(X)$. 
It follows that:
\begin{equation}
- \la \Xi|\ln \Delta_{{\widetilde{\omega}_{1}},\widetilde{\omega}_{\Xi}}|\Xi\ra
= S_{\rm rel}(\rho_1|\omega_1;\mathcal{A}_L).
\end{equation}
Similarly, we can consider the following relative entropy on $\widetilde A_R^-$:
\begin{equation}
S_{\rm rel}(\widetilde{\omega}_{\Xi}|_R \mid \widetilde{\omega}_{2})_R
= S_{\rm rel}(\rho_2|\omega_2;\mathcal{A}_R)
\end{equation}
where we defined $\widetilde{\omega}_{2}(\cdot \otimes 1) = \omega_2(\cdot)$
on $\widetilde \A_R^-
= \mathcal{A}_R \otimes 1$ and:
\begin{equation}
   \rho_2(\cdot) = \int dp |\widetilde g(p)|^2 \omega_{\psi,R} \circ \Delta_{\w_2}^{ip} (\cdot) \Delta_{\w_2}^{-ip}.
\end{equation}
Therefore, we find that the difference of relative entropies
\begin{equation}
\begin{aligned}
    S_{\rm rel}(\widetilde{\omega}_{\Xi}|{\widetilde{\omega}_{T}})_{MR} - S_{\rm rel}(\widetilde{\omega}_{\Xi}|_R|\widetilde{\omega}_{2})_{R} &=S_{\rm rel}(\rho_1|\omega_1;\mathcal{A}_L)-S_{\rm rel}(\rho_2|\omega_2;\mathcal{A}_R) -\expval{(X_--h_-)}{\Xi}\\
    &=S_{\rm rel}(\rho_1|\omega_1;\mathcal{A}_L)-S_{\rm rel}(\rho_2|\omega_2;\mathcal{A}_R) -\expval{X_-}{\widehat{\Xi}}
    \label{exactlyfind}
\end{aligned}
\end{equation}
where $|\widehat \Xi \ra$ is the classical-quantum state without the twirling factor. 
Note that the latter expectation
is simply the average
of $\left< X_- \right>_{|g|^2}$
using the classical distribution $|g|^2$.

The above expression is exact, and already resembles the generalized entropy difference computed using the outside algebras. To make this more clear we consider a semiclassical
like limit where $\widetilde g(p)$ is sharply peaked at $p=0$. The state $\rho_1,\rho_2$ are approximately $\omega_{\psi,L}$ and $\omega_{\psi,R}$, the states induced by $|\psi\ra$ on $\A_L$ and $\A_R$. Thus the relative entropies can be approximated by those of $\psi$:
\begin{equation}
\begin{aligned}\label{petz like formula}
    S_{\rm rel}(\widetilde{\omega}_{\Xi}|{\widetilde{\omega}_{T}})_{MR} - S_{\rm rel}(\widetilde{\omega}_{\Xi}|\widetilde{\omega}_{2})_{R} 
    &\approx S_{\rm rel}(\omega_\psi|\w_1;\mathcal{A}_L)-S_{\rm rel}(\omega_\psi|\w_2;\mathcal{A}_R) -\expval*{X_-}{\hat \Xi}\\
    &=S_{\rm rel}(\omega_\psi|\w_2\circ T;\mathcal{A}_{MR})-S_{\rm rel}(\omega_\psi|\w_2;\mathcal{A}_R)-\expval*{X_-+h_-}{\hat \Xi}.
\end{aligned}
\end{equation}
To obtain the last line we again used cocycle relations as above,  but now for the QFT algebras. This allows us to trade the relative entropy on $\A_L$ for that on $\A_{MR}$ at the expense
of reintroducing $\left< \psi\right| h_i  \left| \psi \right>$.

The above expression can now be directly shown to equal the generalized entropy difference between regions $MR$ and $R$:
\begin{equation}
   \left( \frac{A_1}{4G} + S_L(\psi)\right) -\left( \frac{A_2}{4G}+S_R(\psi)\right).
\end{equation}
To see this, we formally write all quantum field theory states in terms of density matrices. Note that the weight $\w_2\circ T$ on $\A_{MR}$ can be formally written as a factorized density matrix $\bm\rho_{\w_2}^R\tp\bm\s$ where $\bm\s$ acts on the middle region and is uniquely determined by $T$, we then have 
\begin{align}
    S_{\rm rel}(\omega_\psi|\w_2\circ T;\A_{MR})-S_{\rm rel}(\omega_\psi|\w_2;\A_R)=&\tr_{MR}(\bm\rho^{MR}_\psi\ln\bm\rho^{MR}_\psi)-\tr_{MR}(\bm\rho^{MR}_\psi\ln\bm\rho^{R}_{\w_2})-\tr_{MR}(\bm\rho^{MR}_\psi\ln \bm\s)\nonumber\\
    &-\tr_R\big[(\tr_M\bm\rho^{MR}_\psi)\ln(\tr_M\bm\rho^{MR}_\psi)\big]+\tr_R\big[
    (\tr_M\bm\rho^{MR}_\psi)\ln \bm\rho_{\w_2}^R\big]\nonumber\\
    =&-S_\psi(\A_{MR})+S_\psi(\A_R)-\expval{\ln\bm\s}_\psi
\end{align}
where we used the fact that $\tr_{MR}(\bm\rho^{MR}_\psi\ln\bm\rho^{R}_{\w_2})=\tr_R\big[
(\tr_M\bm\rho^{MR}_\psi)\ln \bm\rho_{\w_2}^R\big]$. Using~\eqref{reg diff area} we get
\begin{equation}
    \expval{X_-+h_-}_{\hat\Xi}-S_{\rm rel}(\omega_\psi|\w_2\circ T;\mathcal{A}_{MR})-S_{\rm rel}(\omega_\psi|\w_2;\mathcal{A}_R)=S_{\rm gen}(\A_{MR})-S_{\rm gen}(\A_R)
\end{equation}
which is exactly~\eqref{petz like formula}. 

Interestingly, this quantity computed from the difference in relative entropies does not have the term from the area fluctuation $\int dX_- |g(X_-)|^2 \ln |g(X_-)|^2$. However, this is expected since our classical-quantum state is a pure state on the $B(L^2(\mathbb{R}))$ factor of $\widetilde \A_{MR}^-$, therefore we have zero entanglement entropy from fluctuation of $X_-$. It is also not possible to get the entropy term from fluctuation of $X_+$\footnote{As we have discussed in the beginning of this section, $X_+$ lives in a factorized center in this case. Therefore the fluctuation of $X_+$ contributes non-trivially in the entropy calculation.}, since $X_+$ is the center of both $\widetilde \A_{MR}^-$ and $\widetilde \A_R^-$, and entropy from $X_+$ should cancel out when taking the difference.

Expression~\eqref{petz like formula} can be viewed as an analog of the conditional entropy. Note that we can write this difference naturally in terms of the operator-valued weight $\til T$ from $\til \A_{MR}^-$ to $\til \A_R^-$ as:
\begin{equation}
  -H(M|R)_{\widetilde{\omega}_{\Xi}} \equiv  S_{\rm rel}(\widetilde{\omega}_{\Xi}|\widetilde{\omega}_{2} \circ \widetilde{T})_{MR} - S_{\rm rel}(\widetilde{\omega}_{\Xi}|\widetilde{\omega}_{2})_{R} 
    = S_{\rm rel}(\widetilde{\omega}_{\Xi}|\widetilde{\omega}_{\Xi}\circ \widetilde{T})_{MR}
    \label{condent}
\end{equation}
where we have applied the well known Petz formula for conditional expectations \cite{ohya2004quantum}. While we are not aware of a proof in the operator-valued weight case, we expect this formula to still apply, perhaps under some restrictions that demand the existence of these relative entropies, see \cite{Longo:2022lod}.\footnote{If we simply use non-normalized conditional expectations the formula still applies. This is suggestive that it might extend to the operator-valued weight case. } In any case we can also explicitly compute the right hand side, using similar tricks to the above computation, and we exactly find \eqref{exactlyfind}: note that
\begin{equation}
\widetilde{\omega}_{\Xi} \circ \widetilde{T}
= \rho_{2}\circ T(\cdot) \otimes {\rm Tr} (e^{X_-}\cdot).
\end{equation}
Using the same trick as above to write $ S_{\rm rel}(\widetilde{\omega}_{\Xi}|\widetilde{\omega}_{\Xi}\circ \widetilde{T})_{MR}$ in terms of the relative entropy on $L$:
\begin{equation}
\begin{aligned}
     S_{\rm rel}(\widetilde{\omega}_{\Xi}|\widetilde{\omega}_{\Xi}\circ \widetilde{T})_{MR}
     &= S_{\rm rel}(\rho_1 | \omega_1;\mathcal{A}_L)
     +\left< \Xi \right| \ln \Delta_{{\widetilde{\omega}_{1}}, \widetilde{\omega}_{\Xi} \circ \widetilde{T}} \left| \Xi \right> \\
     & = S_{\rm rel}(\rho_1 | \omega_1;\mathcal{A}_L)-\left< X_- \right>_g
     + \int dp | \widetilde{g}(p)|^2 \left< \psi \right|
     e^{i p h_-} ( \ln \Delta_{\omega_1, \rho_2 \circ T} -  \ln \Delta_{\omega_1, \omega_2 \circ T} ) e^{-i p h_-}\left| \psi \right>
     \end{aligned}
\end{equation}
but the last term above can be expressed as follows:
\begin{equation}
     =\int dp | \widetilde{g}(p)|^2 \left< \psi \right|
     e^{i p h_-} ( \ln \Delta_{\omega_1 \circ T^{-1}, \rho_2} -  \ln \Delta_{\omega_1 \circ T^{-1}, \omega_2} ) e^{-i p h_-}\left| \psi \right>
\end{equation}
which is then related
to the derivative of the cocycle on $R$ evaluated in the state $\rho_2$: $i\frac{d}{ds}\rho_2( ( \rho_2 : \omega_2)_s )$ at $s=0$, thus reproducing the relative entropy term and Eq. \eqref{exactlyfind}.

It is well known that conditional entropies can be expressed in this form \cite{gao2020relative}, at least in the presence of a maximal conditional expectation -- that is a conditional expectation that restricts to a tracial state on the relative commutant. Recall that $\widetilde{T}$ is tracial when restricted to the type II$_\infty$ gravitational algebra associated to $M$. This gravitational algebra is not the full relative commutant for $\widetilde{T}$, which is instead $\mathcal{A}_M 
\otimes \mathcal{B}(L^2(\mathbb{R}))$.
Since this relative commutant is a type III$_1$ factor there is no tracial weight, instead the best we might hope is to find a semifinite subalgebra that is maximal in the sense that the relative commutant for $\widetilde{\mathcal{A}}^-_M \subset \tilde{\mathcal{A}}_M^-
\otimes \mathcal{B}(L^2(\mathbb{R}))$ is trivial (working in the twirled representation for the former). This is indeed the case. We leave it to future work to figure out if this condition characterizes $\widetilde{T}$, say up to unitary equivalence.

We find it encouraging that  the generalized entropy difference can be written as such a conditional entropy, since conditional entropies are used to characterize the very existence of an entanglement wedge \cite{Akers:2020pmf}. To make this connection precise we need to understand the overall additive constant for this generalized entropy difference. Unfortunately this is not obviously fixed by these considerations even for the difference of entropies. Firstly notice that there are two kinds of normalization ambiguities in our expression for the generalized entropy difference: a) the overall normalization of $T$ is ambiguous for the obvious reason and b) the origin of $X_-$ can be shifted. Since only the combination $T e^{X_-}$ appears in the above expressions,  $T \rightarrow \lambda T, X_- \rightarrow X_- - \ln \lambda$ has no effect.
So, for example, we can use this former freedom to fix some
non-canonical normalization of $T$,  after which we are left only with the ambiguities associated to shifts of $X_-$. 

At the same time if the area of the two extremal surfaces is macroscopically different then our gravitational algebras can only possibly be tracking the fluctuations of the area difference, since the true area difference diverges as $1/G_N$ and so requires some microscopic input to determine. For example the error correction approach to holography \cite{Harlow:2016vwg} fixes such a constant terms. For the asymptotically isometric codes defined in \cite{Faulkner:2022ada}, this ambiguity would be fixed by properties of the code and would determine a sequence of diverging numbers representing $\Delta A/G_N$. In this case the best we can hope is that \eqref{condent} represents corrections to this difference, the ambiguity in $X_-$ can then be passed back and forth between these two terms so that the total is unambiguous. 
On the other hand if we tune the macroscopic part of the area difference to zero then presumably we can fix the remaining ambiguity in $X_-$ by demanding the transition point between the two possible entanglement wedges is delineated by the sign of \eqref{condent}. Of course this is exactly the situation where only state specific entanglement wedges \cite{Akers:2020pmf} can be defined, and the Haag dual criterion of \cite{Faulkner:2022ada} breaks down. However nice states (the compressible states discussed in \cite{Akers:2020pmf}) can be designed to probe the transition precisely and fix the shift ambiguity in $X_-$.
Of course understanding this transition in the first place requires some microscopic input, so the two situations discussed above are similar.

There seems to be at least one case where the normalization of $T$ and origin of $X_-$ are unambiguously defined with minimal reference to the microscopic theory. Consider a symmetry $\mathcal{J}$ (represented as a possibly antiunitary operator on $\mathcal{H}_{QFT}$) that exchanges the two surfaces and corresponding algebras $L \leftrightarrow R$. Then we can look for $T$ which is consistent with this symmetry, such that: $T^{-1} = {\rm Ad}_{\mathcal{J}} \circ T \circ {\rm Ad}_{\mathcal{J}} $ which does fix the normalization of $T$. For example this symmetry exists for the long wormhole with equal temperatures where we use the split construction of $T$ with the same two Hartle-Hawking black hole states on $L$ and $R$. This symmetry extends to the gravitational algebra if we demand it sends $X_- \rightarrow - X_-$. Hence the shift ambiguity is also fixed. 
As long as the code respects this symmetry then \eqref{condent} will directly work as an order parameter without any offset. 

We also note that one-shot holography ``order parameters'' \cite{Akers:2023fqr} may also be defined in our approach.
These will correspond to the generalized min and max conditional entropies. We can define these using well known formulas for the smooth min and max conditional entropies \cite{Koenig:2009avh}, but now generalized to operator-valued weights and applied to the gravitational algebras discussed above:
\begin{equation}
    H_{\min}(M|R)_{\widetilde{\omega}_{\Xi}} 
    = - \inf \{ \ln \Sigma(1) : \widetilde{\omega}_{\Xi} \leq  \Sigma \circ \widetilde{T}, \Sigma \in (\widetilde{\mathcal{A}}^-_{R})_\star^+ \} 
\end{equation}
and $-H_{\max}(M|R)_{\widetilde{\omega}_{\Xi}}$ can be
defined using the same expression for the dual inclusion with $\widetilde{T}$ replaced by $\widetilde{T}^{-1}$.
It is not hard to see that $H_{\rm min}(M|R)_{\widetilde{\omega}_{\Xi}} \leq H(M|R)_{\widetilde{\omega}_{\Xi}} \leq H_{\rm max}(M|R)_{\widetilde{\omega}_{\Xi}}$. The aforementioned compressible states are ones where these quantities are sufficiently close.

\section{Discussion}
\label{sec:discussion}
In this paper, we study gravitational algebras on spacetimes with two extremal surfaces, focusing specifically on the assignment of gravitational algebras to various regions in a long wormhole. We discuss three different microcanonical ensembles in the $G_N\to 0$ limit. In one case, we have microcanonical fluctuations for both $H_L$ and $H_R$, where we make use of the split property to construct type II algebras, whose entropies match with the corresponding generalized entropies. For the microcanonical ensemble in only $H_L-H_R$, we make use of operator-valued weights to construct the gravitational algebra. We show that differences in generalized entropies for two different regions can be obtained from differences of the relative entropies for the two gravitational algebras.

Despite the progress we have made in this paper, more work is needed to connect our results to related topics and more general setups. Hereby we list some directions to pursue the in future.  

\subsection*{Quantum field theory on kinked Cauchy slices}

In Section~\ref{subsec:generalextremal} we have discussed that, in order to deal with extremal surfaces in general backgrounds, we need to consider evolutions of Cauchy slices with kinks, which include initial data with delta functions in the extrinsic curvature. In our current work and previous literature, we always assume a vanishing shear for the extremal surface, which is guaranteed by spherical symmetry. In more general cases, this assumption could fail. In the presence of extremal surfaces with non-vanishing shear, ADM evolution of a kinked Cauchy slice leads to  Weyl tensor shocks \cite{Bousso:2020yxi}. Therefore we need to study quantum field theories on these singular geometries to define Hilbert spaces $\H_T$ as discussed in Section~\ref{subsec:generalextremal}. Also, it remains unclear how to construct the intertwiner $V_T$ and the representations of von Neumann algebras $\pi_T$ and $\pi'_T$ in a general setup, which we will leave to future works.

\subsection*{Bulges and multiple extremal surfaces}
In this paper, we only consider the area operators of the two locally minimal extremal surfaces. However, there is another extremal surface between them, which is a local maximum in area. This is sometimes referred to as the ``bulge'' surface. One could consider regions with the bulge surface as part of the boundary, and ask what the gravitational algebra should be. The operator that consists of the bulge area and an appropriate modular Hamiltonian seems to be a natural choice to include in the algebra. However, this can be subtle since the bulge surface can sometimes spontaneously break the symmetry of the spacetime background \cite{Arora:2024edk}. 

We can also ask what the gravitational algebra looks like in the presence of multiple extremal surfaces, which should be a direct generalization of the current setup, where one might need to implement the split property consecutively and multiple operator-valued weights that are nested to construct desired gravitational algebras. Another interesting generalization of our construction is the case where there are regions bounded by more than two extremal surfaces, for example, multi-boundary wormholes in AdS$_3$. 

\subsection*{Enhanced entropy corrections near phase transitions}
When there are two competing quantum extremal surfaces, there is an $O(1/\sqrt{G_N})$ enhanced correction to the holographic entanglement entropy \cite{Marolf:2020vsi,Dong:2020iod,Penington:2019kki,Akers:2020pmf}. This correction has been derived  from the sum over multiple bulk saddles, including those breaking the replica symmetry, in the gravitational path integral. Since in this work we studies the gravitational algebras on such spacetimes, it is natural to ask whether this enhancement of entropy corrections manifests itself in the gravitational algebras.

Before proceeding, we should notice that the algebras we discuss are purely from the bulk, and so are the density matrices and entropies. On the other hand, the enhanced corrections are for entanglement entropy in the boundary CFT. Therefore, the first question to address is to understand what gravitational algebras are dual to the operator algebra for a CFT subregion, when we are near a phase transition. After this, we can ask if it is possible to derive the ``diagonal approximation'' \cite{Marolf:2020vsi,Penington:2024jmt} for the density matrix in the two areas $A_1, A_2$ which is crucial for deriving the entropy correction.

\subsection*{Boundary-anchored extremal surfaces and operator-valued weights}
The application of the split property in our long wormhole example depends on the existence of a `gap' region between the two extremal surfaces. However, this is not guaranteed in arbitrary geometries with competing extremal surfaces. For example, in Figure~\ref{fig:PL-Vacuum AdS} the two extremal surfaces (red and blue curves) touch on the asymptotic boundary. In this case, due to the IR divergence of areas, only the area difference is well-defined, which needs to be defined through an operator-valued weight as discussed in Section~\ref{sec:5.2}. It seems that the split would fail in this case. Nevertheless, as we have shown in Section~\ref{sec:3.4}, split property is a sufficient but not necessary condition for the existence of operator-valued weights between nested subregions. So it remains unclear whether such a operator-valued weight exist in this case. 

Moreover, in Section~\ref{sec:3.4} we only proved the existence of operator-valued weights in the case that nested subregions share a whole boundary, and whether the conclusion remains true if they only share a part of the boundary is not yet fully understood. In future, we plan to elaborate more on the existence of operator-valued weights in quantum field theories. Our goal is to find an operational way to determine whether operator-valued weights exist in given setups. Namely, we want to identify field theoretic quantities which serve as criteria for the existence of operator-valued weights and can be computed at least in some simple cases. Similar criteria have been proposed for the split property~\cite{Buchholz:1973bk,Summers:1982np,Buchholz:1985ii,Buchholz:1986dy,Werner:1987rw}, and we expect them to exist for operator-valued weights, possibly by losing some conditions which lead to split property.

\subsection*{Operator-valued weight, complexity and non-isometric encoding}
In~\cite{Brown:2019rox} the complexity of decoding python's lunch from its local (not global) minimal surface end is given by $\C\propto \exp(\frac{1}{2}\frac{A_B-A_R}{4G})$, where $A_B$ is the area of the bulge and $A_R$ is the area of the local minimal surface (which we assume to be the one on the right). This in turn gives the complexity of bulk reconstruction within the python's lunch region. It is then natural to ask whether this complexity can be accounted for within the algebraic framework we constructed. One subtlety is that from the above expression the complexity necessarily involves knowledge of the bulge surface, which remains elusive in our approach. So understanding complexity requires us to incorporate the bulge surface into the gravitational algebra. Another question is which part in our construction could encode the complexity. One natural guess is that operator-valued weights between gravitational algebras of different subregions potentially keep track of the complexity. 
In particular the construction of $T$ using the split involves two ingredients, an isometric map $E_1$ followed by an operator-valued weight, see around \eqref{condexp}. This looks somewhat similar to the tensor network picture of the python's lunch geometry \cite{Brown:2019rox}, where there is an isometric portion of the network and a portion of the network where postselection is required. Using this analogy one is tempted to replace the bulge surface with the the type-$I$ splitting algebra - this is no longer an algebra associated to geometric region, since that would necessarily be type-III$_1$, but there is no obvious reason not to include such algebras when looking for complexity bottlenecks. 

\begin{figure}[h!]
  \centering
  \includegraphics[width=0.7\textwidth]{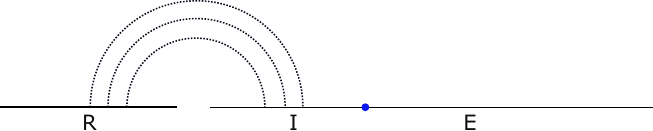} 
  \caption{An intuitive illustration of an evaporating black hole after Page time, where $I$, $E$, $R$ respectively denote the interior, exterior and the radiation respectively.} 
  \label{evaporating blackhole}
\end{figure}

One other related question is regarding an evaporating black hole, as shown in Figure~\ref{evaporating blackhole}. In this case following a similar proof as in Section~\ref{sec:3.4}, we can show that there is an operator-valued weight from the union $I\cup E$ to $E$ where $I$ and $E$ are respectively the interior and exterior of the black hole. In this case our proposal is that the operator-valued weight may contain some information about the non-isometric encoding of the effective description into the fundamental description discussed in~\cite{Akers:2022qdl}. We expect that the asymptotically isometric codes in \cite{Faulkner:2022ada}, which are intrinsically non-isometric yet can behave very similar to isometric codes, will become ``more'' non-isometrical in the presence of the bulk algebras and operator-valued weight discussed in this paper.  
We will leave the verification of these proposals to our future works.   

\acknowledgments
We would like to thank Yasin Alam, Ping Gao, Marius Junge, Marc Klinger, Sarah Racz, Gautam Satishchandran, Zixia Wei, Jiuci Xu for valuable discussions. This work is supported by the DOE award number DE-SC0015655.

\appendix
\section{Crossed product from the boundary perspective}
\label{app:a}
In this appendix we review Witten's construction in~\cite{Witten:2021unn}, which can be understood as the boundary dual picture of the construction in Section~\ref{sec:2.1}. Consider the boundary thermofield double state which is dual to a two-sided AdS black hole in the bulk
\begin{equation}
    \ket{\Psi_{\text{TFD}}}=\frac{1}{\sqrt{Z}}\sum_i e^{-\beta E_i/2}\ket{E_{iL}}\otimes\ket{E_{iR}}
\end{equation}
where $\beta$ is the inverse temperature of the black hole. Despite the explicit $\b$ dependence, this state also depends implicitly on the large parameter $N$. Now consider the algebra of single-trace operators on left/right boundaries denoted by $\mathcal A^{(N)}_{L/R}$. GNS construction with respect to $\ket{\Psi_{\text{TFD}}}$ leads to a Hilbert space $\mathcal H_{\text{GNS}}^{(N)}$ for each $N$. We can also take the large $N$ limit by considering suitable large $N$ limits of correlation functions of single trace operators. In this limit we have a Hilbert space which can be interpreted as the quantum field theory Hilbert space on a fixed two-sided AdS black hole geometry\footnote{We can include small (order $\sqrt {G_N}$) fluctuations of the metric, but they are treated as a quantum field on a fixed background.}. The von Neumann algebras generated by single trace operators in the left and right boundary CFT are dual to the QFT algebras in the left and right exteriors, and are thus type III$_1$ factors \cite{Leutheusser:2021qhd,Leutheusser:2021frk}. We denote them as $\A_L$ and $\A_R$ following the notation in Section~\ref{sec:2.1}. The relation between bulk QFT operators and boundary single trace operators is encoded in the HKLL reconstruction map \cite{Hamilton:2005ju,Hamilton:2006az}.

One important observation is that $\A_L$ and $\A_R$ do not contain boundary Hamiltonians. Let $H_L$ and $H_R$ be the boundary Hamiltonians for the left and right CFTs. They are dual to the left and right ADM Hamiltonians in the bulk and satisfy the constraint $H_L=H_R$ in the large $N$ limit. Naively we may expect the following subtracted boundary Hamiltonian to be an element of $\A_R$
\begin{equation}
    H'_R=H_R-\expval{H_R}
\end{equation}
Unfortunately $H_R'$ is in fact not an operator, although it has an vanishing average. From the form of $\ket{\Ps_{\text{TFD}}}$ the fluctuation of $H'_R$ can be easily seen to be of order $N^2$, which diverges in the large-$N$ limit. One could then ask if the following renormalized operator works 
\begin{equation}
    U=\frac{H_R'}{N}
\end{equation}
This is a legitimate operator as it has finite fluctuation. However, if we add this operator to $\A_R$, then it lives in the center of the resulting algebra\footnote{The center of an algebra is defined to be the set of all elements which commutes with every element in the algebra.}. This can be verified by noticing the fact that for arbitrary $a\in\A_R$, we have 
\begin{equation}\label{U commutator}
    [U,a]=-\frac{i}{N}\partial_t a\rightarrow 0.
\end{equation}
Therefore we obtain a non-factor algebra in the strict large-$N$ limit by adding $U$. The Hilbert space on which this algebra acts is given by
\begin{equation}
    \mathcal H=\mathcal H_{QFT}\otimes L^2(\mathbb R)
\end{equation}
where $U$ acts on the $L^2(\mathbb R)$ factor. In this case the operator $U$ and $\A$ does not talk to each other, which is equivalent to the fact that as $G_N\sim\frac{1}{N^2}$, taking the strict large-$N$ limit amounts to turn off the gravity in the bulk. In this case there is no backreaction on the background geometry from matter. 

Now we turn on the $\frac{1}{N}$ perturbation, at this level $\frac{H_R'}{N}$ and $\mathcal A$ cease to commute, instead their commutator is given by
\begin{equation}\label{cross product commutator}
    [\frac{H_R'}{N},a]=-\frac{i}{N}\partial_t a
\end{equation}
Therefore the identification $\frac{H_R'}{N}=U$ no longer holds. But notice that we can satisfy this commutation relation my making the identification
\begin{equation}\label{rescaled right hamiltonian}
    \frac{H_R'}{N}=U+\frac{\hat h}{N}
\end{equation}
where $\hat h$ is the two-sided boost operator in the bulk, which is the modular Hamiltonian induced by the bulk thermofield double state. So we conclude that in this case the algebra in the right exterior is generated by $\mathcal A_{R}$ and $U+\frac{\hat h}{N}$, which is a crossed product algebra.
\begin{equation}
    \widetilde\A_R=\mathcal A_{R}\rtimes_{\sigma}\mathbb R
\end{equation}
The commutant of $\widetilde\A_R$ in this case is
\begin{equation}
    \widetilde \A_L=\widetilde\A_R'=\{e^{\frac{i\Pi\hat h}{N}}\mathcal A_{L}e^{-\frac{i\Pi \hat h}{N}},U\}''
\end{equation}
where $\Pi$ is the momentum operator conjugate to $U$:
\begin{equation}
    [U,\Pi]=i
\end{equation}
An important generalization of the construction above is the microcanonical ensemble case \cite{Chandrasekaran:2022eqq}. Instead of the state $\ket{\Ps_{\text{TFD}}}$ we can start from the following microcanonical thermofield double state
\begin{equation}
    \ket{\Psi_{\text{MC}}}=\frac{1}{\sqrt{Z}}\sum_i f(E_i)\ket{E_{iL}}\otimes\ket{E_{iR}}
\end{equation}
where $f(E_i)$ is a function peaked around the average energy of the system corresponding to a inverse temperature $\beta$. From statistical physics we know that in the large $N$ limit observables acquire identical expectation values in both ensembles. Typically, they lead to identical correlation functions for single trace operators and therefore have identical bulk duals. The key difference is that the fluctuation of $H_R'$ is of different order in these two cases. With the microcanonical state we have $\expval{(H_R')^2}\sim\mathcal O(1)$. Therefore we can directly include the operator $H'_R$ in the algebra without the $\frac{1}{N}$ rescaling. The commutator~\eqref{cross product commutator} is replaced by 
\begin{equation}\label{cross product commutator 2}
    [H_R',a]=-i\partial_t a
\end{equation}
And in this case~\eqref{rescaled right hamiltonian} becomes
\begin{equation}
    H_R'=U+\hat h
\end{equation}
where $\hat h$ is the same boost generator as in the canonical case. The right algebra is
\begin{equation}
    \widetilde \A_R=\mathcal A_{R}\rtimes_{\sigma}\mathbb R
\end{equation}
and the left algebra is 
\begin{equation}
    \widetilde \A_L=\widetilde\A_R'=\{e^{i\Pi\hat h}\mathcal A_{L}e^{-i\Pi \hat h},U\}''
\end{equation}
As before $\widetilde \A_{L,R}$ are both type $\text{II}_\infty$ algebras in this case.

The difference between canonical and microcanonical ensembles has the bulk interpretation. As we discussed in Section~\ref{sec:2.1}, the $H_R$ (or equivalently $H_R'$) is dual to the bulk time shift operator $t$, and they satisfy the commutation relation $[H_R,t]\sim i$. Therefore the fluctuation of $H_R'$ determines the fluctuation of $t$ by uncertainty principle. In the canonical case the fluctuation of $H_R'$ is of order $N\sim \frac{1}{\sqrt{G_N}}$, which in turn leads to an order $\sqrt{G_N}$ fluctuation of $t$. In contrast, fluctuations of both $H_R'$ and $t$ are of order $1$ in the microcanonical case. That is to say, going to the microcanonical ensemble amounts to enlarge the fluctuation of the bulk geometry by a factor $\frac{1}{\sqrt{G_N}}$.

\section{Trace and density matrices in the crossed product algebra}
\label{app:b}
In this subsection we review some key results for crossed product algebras. For more details see~\cite{takesaki2003theory}. We will focus on the trace and the density matrix, which we use repeatedly in this paper.

To see that~\eqref{trace} is indeed the trace, we only need to verify cyclicity. For two arbitrary operators $\hat a=\int ds\;a(s)e^{is(X+h)}$ and $\hat b =\int ds\;b(s)e^{is(X+h)}$. We have
\begin{align}
    \expval{\hat a\hat b}{\t}&=\int dXdX'dsds'\delta(X-X')e^{isX+is'X'+\frac{X+X'}{2}}\expval{a(s)e^{ish}b(s')}{\eta}\nonumber\\
    &=\int dXdsds'e^{i(s+s')X+X} \expval{a(s)e^{ish}b(s')}{\eta}\nonumber\\
    &=\int dXdsds'e^{i(s+s')X} \expval{a(s)e^{ish}b(s'+i)}{\eta}\nonumber\\
    &=\int ds\expval{a(s)e^{ish}b(-s+i)}{\eta}
\end{align}
Here we used the fact that $h\ket{\eta}=0$ in the first line. We shifted the integration contour $s'\rightarrow s'+i$ to go from the second to the third line, assuming that the function in the integral has nice analytical properties. On the other hand we have 
\begin{align}
    \expval{\hat a\hat b}{\t}&=\int dXdsds'e^{i(s+s')X+X} \expval{a(s)e^{ish}b(s')}{\eta}\nonumber\\
    &=\int dXdsds'e^{i(s+s')X+X} \expval{b(s')e^{-i(s-i)h}a(s)}{\eta}\nonumber\\
    &=\int dXdsds'e^{i(s+s')X} \expval{b(s')e^{-ish}a(s+i)}{\eta}\nonumber\\
    &=\int ds\expval{b(s)e^{ish}a(-s+i)}{\eta}\nonumber\\
    &=\expval{\hat b\hat a}{\tau}
\end{align}
Here in the first line we used the KMS condition, and again shifted $s\rightarrow s+i$ to go from the second to the third line. Therefore we have proved the cyclicity of the trace.

Next we show that~\eqref{density} is indeed the density matrix in the semiclassical limit. Here we only do an approximate calculation up to $O(\e)$ corrections. See \cite{Jensen:2023yxy} for a complete treatment\footnote{Note that the calculation in \cite{Jensen:2023yxy} is done in the twirled representation of the crossed product.}. Again take an operator $\hat a=\int ds\;a(s)e^{is(X+h)}$, we have 
\begin{align}
    \expval{\r_\Psi\hat a}{\t}&=\int dXds \e|g(\e X)|^2e^{isX}\expval{\D_{\psi|\eta}a(s)}{\eta}\nonumber\\
    &=\int dXds \e|g(\e X)|^2e^{isX}\expval{S^\dagger_{\psi|\eta}S_{\psi|\eta}a(s)}{\eta}\nonumber\\
    &=\int dXds \e|g(\e X)|^2e^{isX}\mel{\eta}{S^\dagger_{\psi|\eta}a^\dagger(s)}{\psi}\nonumber\\
    &=\int dXds \e|g(\e X)|^2e^{isX}\expval{a(s)}{\ps}\nonumber\\
    &\sim \int dXds \e|g(\e X)|^2e^{isX}\expval{a(s)e^{ish}}{\ps}+o(\e)
\end{align}
Here we have used the definition of the relative Tomita operator and the relative modular operator, as well as the antilinearity of $S_{\ps|\eta}$ in the first four lines. We have used the fact that whenever $\e |g(\e X)|^2$ is a slowly varying function the integration over $s$ is then supported in the vicinity of $s=0$, therefore inserting $e^{ish}$ factor only causes higher corrections in $\e$.\\ \\

\bibliographystyle{jhep}
\bibliography{biblio}  
\end{document}